\documentclass[aps,prd,reprint,amsmath,amssymb,nofootinbib]{revtex4-2}
\usepackage{mathtools,graphicx,tikz,multirow,pgfplots,makecell}
\usetikzlibrary{fit,shapes.geometric,shapes.misc}
\usepgfplotslibrary{groupplots}
\pgfplotsset{compat=1.3}
\usepackage{newtxtext,newtxmath}
\definecolor{citecolor}{RGB}{45,47,146}
\usepackage[colorlinks,citecolor=citecolor,anchorcolor=red,menucolor=red, linkcolor=citecolor,filecolor=red,runcolor=red,urlcolor=citecolor,frenchlinks=red]{hyperref}

\begin{document}
	\let\oldcite=\cite
	\renewcommand{\cite}[1]{\textcolor{citecolor}{\oldcite{#1}}}
	\renewcommand{\eqref}[1]{\textcolor{citecolor}{(\ref{#1})}}
	
	\title{Investigating three-body resonances in $\alpha\alpha\Omega/\Omega_{ccc}$ clusters within the $\prescript{9}{\Omega/\Omega_{ccc}}{\mathrm{Be}}$ nucleus}
	\author{Hao Zhou$^{1,2,3}$}\email{zhouh2024@lzu.edu.cn}
	\author{Xiang Liu$^{1,2,3,4}$}\email{xiangliu@lzu.edu.cn}
	\affiliation{
		$^1$School of Physical Science and Technology, Lanzhou University, Lanzhou 730000, China\\
		$^2$Lanzhou Center for Theoretical Physics,
		Key Laboratory of Theoretical Physics of Gansu Province,
		Key Laboratory of Quantum Theory and Applications of MoE,
		Gansu Provincial Research Center for Basic Disciplines of Quantum Physics, Lanzhou University, Lanzhou 730000, China\\
		$^3$Research Center for Hadron and CSR Physics, Lanzhou University and Institute of Modern Physics of CAS, Lanzhou 730000, China\\
		$^4$MoE Frontiers Science Center for Rare Isotopes, Lanzhou University, Lanzhou 730000, China}
	\begin{abstract}
We investigate the bound and resonant states of the $\alpha+\alpha+\Omega$ and $\alpha+\alpha+\Omega_{ccc}$ three-body systems, corresponding to $\prescript{9}{\Omega}{\mathrm{Be}}$ and $\prescript{9}{\Omega_{ccc}}{\mathrm{Be}}$, within the Gaussian expansion method combined with the complex scaling method. The $\alpha\Omega$ and $\alpha\Omega_{ccc}$ interactions are constructed by folding the $N\Omega$ and $N\Omega_{ccc}$ potentials obtained from lattice QCD calculations by the HAL QCD Collaboration with the nucleon density distribution of the $\alpha$ particle. The uncertainty associated with the $\alpha$-particle matter radius is also examined. For the $\Omega$ sector, the strong $N\Omega$ attraction generates deeply bound $0_1^+$, $2_1^+$, and $4_1^+$ states in $\prescript{9}{\Omega}{\mathrm{Be}}$, accompanied by a pronounced contraction of the $\alpha\alpha$ core, demonstrating a strong gluelike effect of the $\Omega$ baryon. An unconventional inversion between the $0_1^+$ and $4_1^+$ levels is also predicted. In contrast, the $\Omega_{ccc}$ baryon produces considerably weaker attraction: the $0_1^+$ state of $\prescript{9}{\Omega_{ccc}}{\mathrm{Be}}$ is weakly bound, whereas the $2_1^+$ and $4_1^+$ states remain resonances. Their resonance energies and widths exhibit a clear dependence on the strength of the $\alpha\Omega_{ccc}$ interaction. These results reveal qualitatively different gluelike behaviors of the $\Omega$ and $\Omega_{ccc}$ baryons and provide predictions for the spectroscopy of exotic multistrange and triply charmed hypernuclei.
	\end{abstract}
	
	\maketitle
	
	\section{INTRODUCTION}

The possible existence of dibaryons has attracted sustained theoretical and experimental interest for several decades~\cite{Clement:2016vnl,ExHIC:2017smd}. Among the various candidates, the $N\Omega$ system with strangeness $\mathcal{S}=-3$ and quantum numbers $I(J^{P})=\frac{1}{2}(2^{+})$ is of particular interest. Since the nucleon and the $\Omega$ baryon have no valence quarks of the same flavor, the $N\Omega$ system is free from intercluster quark Pauli blocking and may therefore develop a sizable attractive interaction. Its possible existence as a bound state or a narrow resonance has been extensively investigated using various approaches, including constituent-quark models~\cite{Goldman:1987ma,Oka:1988yq,Silvestre-Brac:1992xsl,Li:1999bc,Li:2000cb,Dai:2007gc,Wang:1995bg,Pang:2004mm,Pang:2003ty,Chen:2011zzb,Zhu:2015sna,Huang:2015yza}, lattice QCD~\cite{HALQCD:2014okw}, chiral effective field theory~\cite{Haidenbauer:2017sws}, QCD sum rules~\cite{Chen:2021hxs}, correlation functions~\cite{Morita:2016auo}, and meson exchanges~\cite{Sekihara:2018tsb}. Motivated by the favorable properties of the $N\Omega$ system and the expected reduction of kinetic energy in systems containing heavy quarks, its heavy-flavor analogue, $N\Omega_{ccc}$, has also been proposed as a promising $N\Omega$-like dibaryon candidate~\cite{Silvestre-Brac:1992xsl,Huang:2019esu}.

Advances in lattice QCD, particularly the time-dependent HAL QCD method, have significantly promoted studies of $N\Omega$-related few-baryon systems. An exploratory calculation at a heavy pion mass of $m_{\pi}\simeq875$ MeV found an attractive $N\Omega({}^{5}S_{2})$ interaction and a relatively deeply bound state with $B\simeq18.9$ MeV~\cite{HALQCD:2014okw}. Near the physical point, $m_{\pi}\simeq146$ MeV, the interaction remained attractive at all distances, whereas the binding was reduced to a shallow bound state near unitarity, with $B=1.54$ MeV from QCD alone~\cite{HALQCD:2018qyu}. The resulting $N\Omega$ potential has subsequently been used as a first-principles two-body input in studies of $NN\Omega$ three-body systems~\cite{Garcilazo:2018gkb,Garcilazo:2019igo,Zhang:2021vsf,Etminan:2023gzh,Filikhin:2025nvt} and $\alpha\Omega$ systems~\cite{Etminan:2019gds,Filikhin:2024hrd,Etminan:2024uvc}. More recently, the $N\Omega_{ccc}$ potentials in the ${}^{3}S_{1}$ and ${}^{5}S_{2}$ channels were extracted at the physical pion mass, $m_{\pi}\simeq137$ MeV. Both interactions were found to be attractive, although neither supports a two-body bound state~\cite{Zhang:2025zaa}, thereby motivating further investigations of possible $NN\Omega_{ccc}$ three-body states~\cite{Etminan:2025emv,Filikhin:2025ige}. 

The systems discussed above contain at most two nucleons and therefore belong to the light few-baryon sector. A natural extension is to embed a hyperon in a many-nucleon environment, thereby forming a hypernucleus. Hypernuclear studies have long focused on $\Lambda$ hypernuclei, with benchmark systems such as ${}^{9}_{\Lambda}\mathrm{Be}$ providing valuable insights into hypernuclear cluster structures and the  ``gluelike" effect of a $\Lambda$ hyperon~\cite{Gal:2016boi,Botta:2012xi,Hiyama:2009zz,Hashimoto:2006aw,Alberico:2001jb,Bando:1990yi,Dover:1985ba}. With increasing strangeness content, the field has gradually expanded from $\Lambda,\Sigma$ to $\Xi$ hypernuclei~\cite{Nakazawa:2015joa,J-PARCE07:2020xbm}, and at a more exploratory level, to $\Omega$ hypernuclei~\cite{Etminan:2022bon,Filikhin:2025ivu}. While substantial experimental and theoretical knowledge has been accumulated for $\Lambda$ hypernuclei and evidence for bound $\Xi$ hypernuclei has emerged, studies of $\Omega$ hypernuclei remain scarce. The discovery and characterization of new hypernuclear species, particularly in the multistrangeness sector, therefore constitute an important frontier in hypernuclear physics. Moreover, hypernuclei provide valuable constraints on hyperon–nucleon and hyperon–nucleon–nucleon interactions, which are essential for understanding hyperonic matter and the equation of state of neutron stars~\cite{Schaffner-Bielich:2008zws,Tolos:2020aln}.

Hypernuclear physics has two complementary objectives. One is to extract information on the underlying hadron--hadron interactions through hypernuclear spectroscopy~\cite{Hiyama:1997ub,Hiyama:2009zz,Hiyama:2012sq}. Once an interaction is specified, the resulting energy spectrum can be confronted with experiment to constrain its detailed properties. A representative example is the determination of the $\Lambda N$ spin--orbit interaction from the spectra of $\prescript{9}{\Lambda}{\mathrm{Be}}$ and $\prescript{13}{\Lambda}{\mathrm{C}}$~\cite{Hiyama:2000jd}. The other objective is to explore new nuclear dynamics induced by an additional non-nucleonic particle~\cite{Hiyama:1996gv}. Since such a particle is not subject to Pauli blocking with the nucleons, it may generate additional bound states and modify, or even contract, the nuclear core, giving rise to the so-called ``gluelike'' effect~\cite{Motoba:1983kbv,Motoba:1985,Bando:1983pv}. 

Recently, these two aspects have been explored in meson-nucleus systems using HAL QCD interactions, including $\phi NN$, $J/\psi NN$, and $\eta_c NN$ systems~\cite{Etminan:2024vkv,Filikhin:2024avj,Wen:2025wit}, as well as $\alpha\alpha M$ nuclei with $M=\phi,J/\psi,\eta_c$~\cite{Filikhin:2024xkb,Etminan:2025yoq,Zhou:2025anp,Wu:2026xrw}. Reference~\cite{Filikhin:2024xkb} employed the Faddeev equations in configuration space for the $\phi+\alpha+\alpha$ and $\phi+\phi+\alpha$ three-cluster systems, predicting bound states with binding energies of 1--11 MeV, while Reference~\cite{Etminan:2025yoq} used the hyperspherical harmonics method for the $\bar{c}c+\alpha+\alpha$ system, obtaining binding energies of about 0.5--1.7 MeV and suggesting a possible Borromean structure. Reference~\cite{Zhou:2025anp} investigated both the spectra and structural changes of the $\alpha\alpha M$ systems, finding a pronounced gluelike effect for the $\phi$ meson, whereas the $J/\psi$ and $\eta_c$ induce much weaker modifications of the nuclear core. Reference~\cite{Wu:2026xrw} performed a rigorous many-body solution of the Schrödinger equation using neural-network variational Monte Carlo for $A=2$ to $12$. In particular, their spectra and structural changes provide a direct testing ground for the lattice-QCD-derived hadron-nucleon potentials. By contrast, analogous many-nucleon systems containing their baryonic counterparts, $\Omega$ and $\Omega_{ccc}$, remain much less explored. Extending these ideas to $\Omega$- and $\Omega_{ccc}$-hypernuclei therefore offers an opportunity to investigate both the corresponding baryon-nucleon interactions and their possible gluelike roles in nuclei.

From a methodological perspective, the Faddeev formalism has been widely used for few-body hypernuclear systems. In the $\Omega$ sector, Reference~\cite{Etminan:2022bon} studied the ground-state properties of the $\Omega\Omega\alpha$ system using the Faddeev equations combined with the hyperspherical-harmonics expansion. More recently, Reference~\cite{Filikhin:2025ivu} extended configuration-space Faddeev calculations to several light $\Omega$- and $\Omega_{ccc}$-containing nuclei, mainly focusing on their bound-state and ground-state properties. Meanwhile, neural-network variational Monte Carlo methods have demonstrated promising capabilities for directly solving many-body Schrödinger equations~\cite{Wu:2026xrw}. Complementary to these approaches, the Gaussian expansion method combined with the complex scaling method employed in our previous work~\cite{Zhou:2025anp} allows us to systematically investigate not only bound states but also resonant and higher-orbital excited states. This capability is particularly useful for exploring the excitation spectra of $\Omega$- and $\Omega_{ccc}$-hypernuclei beyond their ground-state properties.

The Gaussian expansion method (GEM)~\cite{Hiyama:2003cu,Hiyama:2012sma} is a high-precision variational approach for solving the Schr\"odinger equation in few-body systems. It has achieved numerical accuracy comparable to that of the Faddeev--Yakubovsky method for benchmark systems such as $^3$H ($^3$He) and $^4$He, and has been successfully applied to a wide variety of atomic, baryonic, and quark few-body systems. To describe resonant states, the complex scaling method (CSM)~\cite{Aguilar:1971ve,Balslev:1971vb,Simon:1972} provides a powerful framework in which resonances can be treated on the same footing as bound states. In particular, the CSM has proven highly effective for many-body resonances beyond two-body systems~\cite{Aoyama:2006hrz,Myo:2014ypa}. By combining the CSM with the GEM to solve the complex-scaled Schr\"odinger equation, both bound and resonant states of the system can be systematically investigated.

This paper is organized as follows. Section \ref{sec2} presents the theoretical framework for the $\alpha\alpha\Omega$ and $\alpha\alpha\Omega_{ccc}$ three-body systems, including the interactions and numerical methods employed to describe their bound and resonant states. Section \ref{sec3} presents and discusses the corresponding numerical results. Finally, we summarize our main findings and conclusions.
    
	\section{THEORETICAL FRAMEWORK FOR THE $\alpha\alpha\Omega$ and $\alpha\alpha\Omega_{ccc}$ THREE-BODY SYSTEMS}\label{sec2}
	
	\subsection{Hamiltonian}
	Within the $\alpha+\alpha+\Omega/\Omega_{ccc}$ cluster description, the dynamics are governed by the $\alpha\alpha$ and effective $\alpha\Omega/\Omega_{ccc}$ interactions. We employ a phenomenological potential for the $\alpha\alpha$ subsystem, while the $\alpha\Omega/\Omega_{ccc}$ interactions are constructed by folding the lattice-QCD-derived $N\Omega/\Omega_{ccc}$ potentials with the nucleon density distribution of the $\alpha$ particle. The interaction models and their parametrizations are specified below.
	
	\subsubsection{$NN$ and $\alpha\alpha$ interaction}
    For the nucleon-nucleon interaction, we employ a slightly modified Malfliet-Tjon I-III model~\cite{Malfliet:1968tj,Friar:1990zza}, expressed as a sum of attractive and repulsive Yukawa terms:
    \begin{equation}
        V_{NN}(r)=-V_a\frac{\mathrm{e}^{-\mu_ar}}{r}+V_r\frac{\mathrm{e}^{-\mu_rr}}{r},
    \end{equation}
    where $V_a$ and $\mu_a$ ($V_r$ and $\mu_r$) denote the strength and inverse range of the attractive (repulsive) component, respectively. The parameters for the isospin-singlet $(I=0,\,^{3}S_{1})$ and isospin-triplet $(I=1,\,^{1}S_{0})$ channels are taken from Refs.~\cite{Malfliet:1968tj,Friar:1990zza} and listed in Table~\ref{tab:parameters}. 

The $\alpha\alpha$ interaction is commonly described either within the orthogonality condition model (OCM)~\cite{Saito:1969zz} or by a phenomenological potential. In the OCM, the effective $NN$ interaction and the $pp$ Coulomb interaction are folded into the $\alpha$-cluster wave function, while the Pauli-forbidden components between the two $\alpha$ clusters are removed through the OCM projection operator~\cite{Lee:2019mlt,Wu:2019ivs}. In the phenomenological approach adopted here, the $\alpha\alpha$ interaction is decomposed into nuclear and Coulomb contributions,
\begin{equation}
V(r)=V_N(r)+V_C(r).
\end{equation}
A variety of functional forms have been employed for the nuclear component, including the Morse~\cite{Morse:1929zz}, Woods-Saxon~\cite{Darriulat:1965zz}, Ali-Bodmer double-Gaussian~\cite{Ali:1966olw}, Malfliet-Tjon~\cite{Malfliet:1968tj}, double-Hulthén~\cite{Bhoi:2016wge}, and double-exponential~\cite{Awasthi:2023yre} potentials. Different prescriptions have likewise been considered for the Coulomb term, such as the point-Coulomb, atomic Hulthén~\cite{hulthen:1942}, and modified Coulomb~\cite{Herzenberg:1957} forms.

In the present calculation, we employ the phenomenological $\alpha\alpha$ interaction of Ref.~\cite{Ali:1966olw}. Its nuclear component is parametrized by the double-Gaussian form
\begin{equation}
V_N(r)=V_r\mathrm{e}^{-\mu_r^2r^2}
-V_a\mathrm{e}^{-\mu_a^2r^2},
\end{equation}
where $V_r$ ($V_a$) denotes the strength of the repulsive (attractive) term, and $\mu_r$ ($\mu_a$) is the corresponding inverse range. The Coulomb interaction is taken as
\begin{equation}\label{eq:Coulomb part}
V_C(r)=\frac{4\alpha}{r},
\end{equation}
where $\alpha$ is the fine-structure constant.

The parameters adopted for the $\alpha\alpha$ interaction are listed in Table~\ref{tab:parameters}. As discussed in detail in Ref.~\cite{Zhou:2025anp}, different Ali-Bodmer parameter sets can lead to only modest changes in the binding energies while producing appreciable differences in the resonance properties. We therefore adopt the parameter set recommended in Ref.~\cite{Ali:1966olw}, which provides a more reliable description of the $\alpha\alpha$ subsystem for the present study of both bound and resonant states.
    
	\begin{table*}
		\caption{Parameters of the $NN$, $\alpha\alpha$, $N\Omega$, and $N\Omega_{ccc}$ interactions adopted in the present calculations. For the $\alpha\Omega$ interaction, the Woods-Saxon parameters obtained by fitting the folded potentials are given for $R_\alpha=1.84$, 1.70, and 1.56 fm. The hadron masses used throughout the calculations are also included. Numbers in parentheses indicate the uncertainties of the HAL QCD fit parameters. \label{tab:parameters}}
		\begin{ruledtabular}
\begin{tabular}{cccccccccc}
	                           \multicolumn{5}{c}{$NN$ potential \cite{Malfliet:1968tj,Friar:1990zza}}                            & \multicolumn{5}{c}{$N\Omega$ and $N\Omega_{ccc}$ potential \cite{HALQCD:2018qyu,Zhang:2025zaa}} \\ \cline{1-5}\cline{6-10}
	$I,J$ & $V_a(\mathrm{MeV\,fm})$ &  $\mu_a(\mathrm{fm}^{-1})$  & $V_{r}(\mathrm{MeV\,fm})$ &  $\mu_r(\mathrm{fm}^{-1})$  &         Interaction          &   $a_1$ (MeV)   & $b_1$ (fm)  & $a_2$ (MeV)  &     $b_2$ (fm)     \\ \cline{1-5}\cline{6-10}
	 0,1  &         626.885         &            1.55             &          1438.72          &            3.11             &    $V_{N\Omega}^{J=2}(r)$    &  $-$313.0(5.3)  &  81.7(5.4)  &  $-$252(27)  &      0.85(10)      \\
	 1,0  &         513.968         &            1.55             &          1438.72          &            3.11             & $V_{N\Omega_{ccc}}^{J=2}(r)$ &  $-$52.6(2.5)   &  0.110(12)  & $-$60.4(2.5) &     0.612(50)      \\
	                              \multicolumn{5}{c}{$\hbar^2/m_N=41.47\,\mathrm{MeV\ fm^2}$}                               & $V_{N\Omega_{ccc}}^{J=1}(r)$ &  $-$118.0(3.0)  &  0.135(8)   & $-$80.0(3.7) &     0.601(37)      \\
	                          \multicolumn{5}{c}{$\alpha\alpha$ potential \cite{Ali:1966olw}}                            &                              &             \multicolumn{4}{c}{Woods-Saxon potential}             \\ \cline{1-5}\cline{7-10}
	 $l$  &       $V_r$ (MeV)       & $\mu_r\ (\mathrm{fm}^{-1})$ &        $V_a$ (MeV)        & $\mu_a\ (\mathrm{fm}^{-1})$ &                              & $R_\alpha$ (fm) & $V_0$ (MeV) &   $R$ (fm)   &      $c$ (fm)      \\ \cline{1-5}\cline{7-10}
	  0   &           500           &             0.7             &            130            &            0.475            &                              &      1.84       &    85.3     &     1.41     &       0.613        \\
	  2   &           320           &             0.7             &            130            &            0.475            &                              &      1.70       &     100     &     1.33     &       0.583        \\
	  4   &                         &                             &            130            &            0.475            &                              &      1.56       &     118     &     1.25     &       0.553        \\ \hline
	                \multicolumn{10}{c}{$m_\alpha=3727.3794118$ MeV~\cite{Mohr:2024kco}\quad $m_\pi=139.57039$ MeV\quad $m_\Omega=1672.43$ MeV~\cite{ParticleDataGroup:2024}\quad $m_{\Omega_{ccc}}=4793$ MeV~\cite{Zhou:2025fpp}}
\end{tabular}
		\end{ruledtabular}
	\end{table*}

	\subsubsection{$N\Omega$ and $N\Omega_{ccc}$ interaction}
    For the $N\Omega$ and $N\Omega_{ccc}$ interactions, we adopt the potentials derived from $(2+1)$-flavor lattice QCD simulations within the HAL QCD framework~\cite{HALQCD:2018qyu,Zhang:2025zaa}. These potentials are extracted at nearly physical pion masses and provide a first-principles description of the nonperturbative dynamics between nucleons and $\Omega$ ($\Omega_{ccc}$) baryons.

    The $\Omega$ baryon carries spin $3/2$, so the $N\Omega$ system can be in total spin $S=1$ (triplet, ${}^{3}S_{1}$) or $S=2$ (quintet, ${}^{5}S_{2}$) channels. The HAL QCD collaboration has determined the $N\Omega$ potential in the quintet ($J=2$) channel, which is fitted by a functional form incorporating a two-pion exchange tail:
    \begin{equation}
		V_{N\Omega}^{J=2}(r)=a_1\mathrm{e}^{-b_1r^2}+a_{2}(1-\mathrm{e}^{-b_2r^2})\frac{\mathrm{e}^{-2m_\pi r}}{r^2}. \label{eq:VNOmega}
	\end{equation}
    The Gaussian term models the short-range interaction, while the second term captures the long-range two-pion exchange behavior.

    For the $N\Omega_{ccc}$ system, the HAL QCD potentials are available in both the $J=1$ (${}^{3}S_{1}$) and $J=2$ (${}^{5}S_{2}$) channels. They are parameterized by a sum of two Gaussian functions
	\begin{equation}\label{eq:threeGaussian}
		V_{N\Omega_{ccc}}^J(r)=\sum_{i=1}^{2} a_i\mathrm{e}^{-\frac{r^2}{b_i^2}}.
	\end{equation}
	In addition, except for the cases of $J=1$ and $J=2$, the spin-averaged potential $V_{N\Omega_{ccc}}^{\text{av}}(r)$ used in subsequent calculations is defined by Eq.~\eqref{eq:Vave}. All fitted parameters are collected in Table~\ref{tab:parameters}.

	\subsubsection{$\alpha\Omega$ and $\alpha\Omega_{ccc}$ interaction}
    The $\alpha\Omega$ and $\alpha\Omega_{ccc}$ potentials are constructed by folding the $N\Omega$ and $N\Omega_{ccc}$ potentials into the nucleon density distribution of the $\alpha$ particle via the single folding model~\cite{Satchler:1979ni}:
	\begin{equation}\label{eq:alpha-qq}
		V_{\alpha\Omega}^J(\boldsymbol{r})=\int\rho(\boldsymbol{x})V_{N\Omega}^J(\boldsymbol{r}-\boldsymbol{x})\mathrm{d}\boldsymbol{x},
	\end{equation}
	where $\rho(\boldsymbol{x})$ is the nucleon density function of the $\alpha$ particle in the center-of-mass frame, and $\boldsymbol{r}$ is the vector from the center of mass of the $\alpha$ particle to $\Omega$. Following Ref.~\cite{Wang:2023uek}, we adopt a simple Gaussian form for the nucleon density:
	\begin{equation}
		\rho(\boldsymbol{x})=4\left(\frac{1}{\pi a^2}\right)^{\frac{3}{2}}\mathrm{e}^{-\frac{x^2}{a^2}},
	\end{equation}
	where $a=\sqrt{\frac{2}{3}}R_\alpha$, with $R_\alpha$ being the root-mean-square (rms) matter radius of the $\alpha$ particle.

	For the $N\Omega_{ccc}$ potential, which is expressed as a sum of two Gaussian functions~\eqref{eq:threeGaussian}, the folding integral can be evaluated analytically:
	\begin{equation}\label{eq:analytical alpha-qq}
		V_{\alpha\Omega_{ccc}}^J(r)=\sum_{i=1}^{2}\frac{4a_ib_i^3\mathrm{e}^{-\frac{r^2}{a^2+b_i^2}}}{(a^2+b_i^2)^{3/2}}.
	\end{equation}
	For the $N\Omega$ potential in Eq.~\eqref{eq:VNOmega}, however, the two-pion exchange tail $(1-\mathrm{e}^{-b_2r^2})\mathrm{e}^{-2m_\pi r}/r^2$ cannot be integrated analytically. We therefore evaluate Eq.~\eqref{eq:alpha-qq} numerically and fit the resulting $\alpha\Omega$ potential to a Woods-Saxon form~\cite{Dover:1982ng}:
	\begin{equation}\label{eq:WS potential}
		V_{\alpha\Omega}^J(r)=-\frac{V_0}{1+\mathrm{e}^{\frac{r-R}{c}}},
	\end{equation}
	where $V_0$, $R$, and $c$ denote the depth, radius, and surface diffuseness parameters, respectively. The fitted Woods-Saxon parameters for $R_\alpha=1.84$, 1.70, and 1.56~fm are listed in Table~\ref{tab:parameters}.

	There is a certain degree of uncertainty in the experimental rms charge radius of the $\alpha$ particle, with measured values of 1.681(4)~\cite{Sick:2008zza}, 1.6755(28)~\cite{Angeli:2013epw}, and 1.67824(83)~fm~\cite{Krauth:2021foz}. The rms matter radius was recently determined to be $1.70\pm0.14$~fm~\cite{Wang:2023uek}. To assess the sensitivity of our results to this uncertainty, we perform calculations for three representative values: $R_\alpha=1.84$, 1.70, and 1.56~fm.

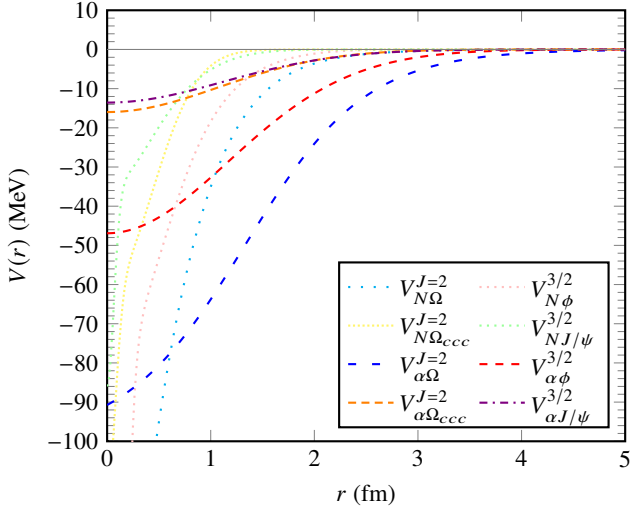
\begin{figure}[htbp]
		\begin{tikzpicture}
			\begin{axis}[
				xlabel=$r$ (fm),
				xmin=0, xmax=5, 
				ylabel=$V(r)$ (MeV),
				ymin=-100, 
				ytick distance=10, minor y tick num=4,
				legend entries={$V_{N\Omega}^{J=2}$, $V_{N\phi}^{3/2}$, $V_{N\Omega_{ccc}}^{J=2}$, $V_{NJ/\psi}^{3/2}$, $V_{\alpha\Omega}^{J=2}$, $V_{\alpha\phi}^{3/2}$, $V_{\alpha\Omega_{ccc}}^{J=2}$, $V_{\alpha J/\psi}^{3/2}$},
                legend cell align=left,
                legend columns=2,
				legend pos=south east,
				thick
				]
                \newcommand\mpp{139.57039*5.067730716156396*1e-3}
				\newcommand\al{-313.0}
				\newcommand\bl{81.7}
				\newcommand\az{-252}
				\newcommand\bz{0.85}
				\addplot+[domain=0.2:5,samples=300,mark=none,loosely dotted,cyan]{\al*exp(-\bl*x*x)+\az*(1-exp(-\bz*x*x))*exp(-2*\mpp*x)/x/x};
                \renewcommand\al{-371}
				\renewcommand\bl{0.15}
				\renewcommand\az{-50}
				\renewcommand\bz{0.66}
				\newcommand\as{-31}
				\newcommand\bs{1.09}
				\addplot+[domain=0.2:5,samples=300,mark=none,dotted,pink]{\al*exp(-x*x/(\bl*\bl))+\az*exp(-x*x/(\bz*\bz))+\as*exp(-x*x/(\bs*\bs))};
                \renewcommand\al{-52.6}
				\renewcommand\bl{0.110}
				\renewcommand\az{-60.4}
				\renewcommand\bz{0.612}
				\addplot+[domain=0:5,samples=300,mark=none,densely dotted,yellow!70]{\al*exp(-x*x/\bl/\bl)+\az*exp(-x*x/\bz/\bz)};
                \renewcommand\al{-51}
				\renewcommand\bl{0.09}
				\renewcommand\az{-13}
				\renewcommand\bz{0.49}
				\renewcommand\as{-22}
				\renewcommand\bs{0.82}
				\addplot+[domain=0:5,samples=300,mark=none,dotted,green!40]{\al*exp(-x*x/(\bl*\bl))+\az*exp(-x*x/(\bz*\bz))+\as*exp(-x*x/(\bs*\bs))};
                
				\newcommand\Vo{100}
                \newcommand\R{1.33}
                \newcommand\cc{0.583}
				\addplot+[domain=0:5,samples=300,mark=none,loosely dashed,blue]{-\Vo/(1+exp((x-\R)/\cc))};
                \renewcommand\al{-371}
				\renewcommand\bl{0.15}
				\renewcommand\az{-50}
				\renewcommand\bz{0.66}
				\renewcommand\as{-31}
				\renewcommand\bs{1.09}
                \newcommand\ax{1.70*sqrt(2/3)}
				\addplot+[domain=0:5,samples=300,mark=none,dashed,red]{4*\al*pow(\bl,3)/pow(\ax*\ax+\bl*\bl,1.5)*exp(-x*x/(\ax*\ax+\bl*\bl))+4*\az*pow(\bz,3)/pow(\ax*\ax+\bz*\bz,1.5)*exp(-x*x/(\ax*\ax+\bz*\bz))+4*\as*pow(\bs,3)/pow(\ax*\ax+\bs*\bs,1.5)*exp(-x*x/(\ax*\ax+\bs*\bs))};
                \renewcommand\al{-52.6}
				\renewcommand\bl{0.110}
				\renewcommand\az{-60.4}
				\renewcommand\bz{0.612}
				\addplot+[domain=0:5,samples=300,mark=none,densely dashed,orange]{4*\al*pow(\bl,3)/pow(\ax*\ax+\bl*\bl,1.5)*exp(-x*x/(\ax*\ax+\bl*\bl))+4*\az*pow(\bz,3)/pow(\ax*\ax+\bz*\bz,1.5)*exp(-x*x/(\ax*\ax+\bz*\bz))};
                \renewcommand\al{-51}
				\renewcommand\bl{0.09}
				\renewcommand\az{-13}
				\renewcommand\bz{0.49}
				\renewcommand\as{-22}
				\renewcommand\bs{0.82}
				\addplot+[domain=0:5,samples=300,mark=none,dash dot,violet]{4*\al*pow(\bl,3)/pow(\ax*\ax+\bl*\bl,1.5)*exp(-x*x/(\ax*\ax+\bl*\bl))+4*\az*pow(\bz,3)/pow(\ax*\ax+\bz*\bz,1.5)*exp(-x*x/(\ax*\ax+\bz*\bz))+4*\as*pow(\bs,3)/pow(\ax*\ax+\bs*\bs,1.5)*exp(-x*x/(\ax*\ax+\bs*\bs))};
                \draw[gray,thin](axis cs:0,0)--(axis cs:5,0);
			\end{axis}
		\end{tikzpicture}
    \caption{Comparison of different HAL QCD potentials. The dotted curves show the $N\Omega$~\cite{HALQCD:2018qyu}, $N\phi$~\cite{Lyu:2022imf}, $N\Omega_{ccc}$~\cite{Zhang:2025zaa}, and $NJ/\psi$~\cite{Lyu:2024ttm} interactions, while the dashed and dash-dotted curves represent the corresponding folded $\alpha\Omega$, $\alpha\phi$, $\alpha\Omega_{ccc}$, and $\alpha J/\psi$ potentials. Their attractive strengths decrease in this order.}\label{fig:HAL-potentials}
 \end{figure}

	Figure~\ref{fig:HAL-potentials} compares the $N\Omega$~\cite{HALQCD:2018qyu}, $N\phi$~\cite{Lyu:2022imf}, $N\Omega_{ccc}$~\cite{Zhang:2025zaa}, and $NJ/\psi$~\cite{Lyu:2024ttm} HAL QCD potentials with the maximal spin together with the corresponding folded $\alpha\Omega$, $\alpha\phi$, $\alpha\Omega_{ccc}$, and $\alpha J/\psi$ potentials calculated from Eq.~\eqref{eq:alpha-qq}. It can be seen that the attractive strength decreases in the order $N\Omega > N\phi > N\Omega_{ccc} > NJ/\psi$, and the same trend is reflected in the folded $\alpha$-baryon/meson potentials. The $N\Omega$ interaction is particularly deep at short distances, while the $N\Omega_{ccc}$ and $NJ/\psi$ interactions are considerably shallower. 

	Figure~\ref{fig:potential-comparison} provides a detailed comparison of different treatments of the $N\Omega$ potential and the resulting $\alpha\Omega$ folded potentials. The pink dotted curve shows the original HAL QCD $N\Omega$ potential in the ${}^5S_2$ channel. The cyan dotted curve represents the modified $N\Omega$ potential from Ref.~\cite{Filikhin:2025ivu}, where the short-range part has been removed. The red solid curve and the blue dashed curve are the $\alpha\Omega$ potentials obtained by folding the original HAL QCD and the modified $N\Omega$ potentials, respectively, using the single folding model with $R_\alpha=1.70$~fm (Eq.~\eqref{eq:alpha-qq}). The black dash-dotted curve is the double Gaussian fit to the long-range ($r\geq2$~fm) part of the $\alpha\Omega$ potential performed in Ref.~\cite{Filikhin:2025ivu}. It is noted that removing the short-range part of the $N\Omega$ HAL QCD potential has little impact on the $\alpha\Omega$ potential obtained using the single-folding model. The actual cause of the shallowing of the $\alpha\Omega$ potential is the double Gaussian fitting in the region of $r\geq2$~fm as described in Ref.~\cite{Filikhin:2025ivu}. In our view, directly employing the original HAL QCD potential without any ad~hoc modification of the short-range part is physically more justified, as the HAL QCD potential is derived from first-principles lattice QCD simulations and contains both short- and long-range physics on an equal footing.

    \subsubsection{Spin-averaged interaction}
    The interaction potentials are inherently spin-dependent: the different coupling schemes between the $\Omega$ baryon and the nucleons directly affect the effective interaction strength between the particles. To determine the proper weighting of different spin channels, we first analyze the spin-isospin configurations of the $NN\Omega$ system, and then extend the discussion to the $\alpha\Omega$ case.

\begin{figure}
        \begin{tabular}{c@{}c}
            \begin{tikzpicture}[N/.style={circle,fill=cyan!40,inner sep=1pt},phi/.style={circle,fill=red!40,inner sep=1pt},ellip/.style={ellipse,draw=gray,line width=0.4pt,dash pattern=on 1pt off 1pt,inner sep=0pt}]
                \node at(0,3){$S=\dfrac{1}{2}$};
                \begin{scope}[shift={(120:2)}]
                    \node(N1)at(120:0.3)[N]{$N$};
                    \node(N2)at(240:0.3)[N,label={[label distance=-2pt]-90:$s=1/2$}]{$N$};
                    \node(phi)at(0:0.3)[phi]{$\Omega$};
                    \node[ellip,rotate fit=60,fit=(N1)(phi),label={[label distance=-3pt]0:$s=1$}]{};
                \end{scope}
                \begin{scope}[shift={(60:2)}]
                    \node(N1)at(120:0.3)[N,label={[label distance=-5pt]60:$s=1/2$}]{$N$};
                    \node(N2)at(240:0.3)[N]{$N$};
                    \node(phi)at(0:0.3)[phi]{$\Omega$};
                    \node[ellip,rotate fit=30,fit=(N2)(phi),label={[label distance=-5pt]-90:$s=1$}]{};
                \end{scope}
                \begin{scope}[shift={(-90:-0.1)}]
                    \node(N1)at(120:0.25)[N]{$N$};
                    \node(N2)at(240:0.25)[N]{$N$};
                    \node(phi)at(0:0.5)[phi,label={[label distance=-2pt]-90:$\hspace{10pt}s=3/2$}]{$\Omega$};
                    \node[ellip,fit=(N1)(N2),label={[label distance=-5pt]80:$s=1,t=0$}]{};
                    \node at(180:0.13){\scriptsize$d$};
                \end{scope}
            \end{tikzpicture}&
            \begin{tikzpicture}[N/.style={circle,fill=cyan!40,inner sep=1pt},phi/.style={circle,fill=red!40,inner sep=1pt},ellip/.style={ellipse,draw=gray,line width=0.4pt,dash pattern=on 1pt off 1pt,inner sep=0pt}]
                \node at(0,3){$S=\dfrac{5}{2}$};
                \begin{scope}[shift={(120:2)}]
                    \node(N1)at(120:0.3)[N]{$N$};
                    \node(N2)at(240:0.3)[N,label={[label distance=-2pt]-90:$s=1/2$}]{$N$};
                    \node(phi)at(0:0.3)[phi]{$\Omega$};
                    \node[ellip,rotate fit=60,fit=(N1)(phi),label={[label distance=-3pt]0:$s=2$}]{};
                \end{scope}
                \begin{scope}[shift={(60:2)}]
                    \node(N1)at(120:0.3)[N,label={[label distance=-5pt]60:$s=1/2$}]{$N$};
                    \node(N2)at(240:0.3)[N]{$N$};
                    \node(phi)at(0:0.3)[phi]{$\Omega$};
                    \node[ellip,rotate fit=30,fit=(N2)(phi),label={[label distance=-5pt]-90:$s=2$}]{};
                \end{scope}
                \begin{scope}[shift={(-90:-0.1)}]
                    \node(N1)at(120:0.25)[N]{$N$};
                    \node(N2)at(240:0.25)[N]{$N$};
                    \node(phi)at(0:0.5)[phi,label={[label distance=-2pt]-90:$\hspace{10pt}s=3/2$}]{$\Omega$};
                    \node[ellip,fit=(N1)(N2),label={[label distance=-5pt]80:$s=1,t=0$}]{};
                    \node at(180:0.13){\scriptsize$d$};
                \end{scope}
            \end{tikzpicture}\\[1.5em]
            \begin{tikzpicture}[N/.style={circle,fill=cyan!40,inner sep=1pt},phi/.style={circle,fill=red!40,inner sep=1pt},ellip/.style={ellipse,draw=gray,line width=0.4pt,dash pattern=on 1pt off 1pt,inner sep=0pt}]
                \node at(0,3){$S=\dfrac{3}{2},T=0$};
                \begin{scope}[shift={(120:2)}]
                    \node(N1)at(120:0.3)[N]{$N$};
                    \node(N2)at(240:0.3)[N,label={[label distance=-2pt]-90:$s=1/2$}]{$N$};
                    \node(phi)at(0:0.3)[phi]{$\Omega$};
                    \node[ellip,rotate fit=60,fit=(N1)(phi),label={[label distance=-3pt]0:$s=1$}]{};
                \end{scope}
                \begin{scope}[shift={(60:2)}]
                    \node(N1)at(120:0.3)[N]{$N$};
                    \node(N2)at(240:0.3)[N,label={[label distance=-2pt]-90:$s=1/2$}]{$N$};
                    \node(phi)at(0:0.3)[phi]{$\Omega$};
                    \node[ellip,rotate fit=60,fit=(N1)(phi),label={[label distance=-3pt]0:$s=2$}]{};
                \end{scope}
                \begin{scope}[shift={(-90:0)}]
                    \node(N1)at(120:0.25)[N]{$N$};
                    \node(N2)at(240:0.25)[N]{$N$};
                    \node(phi)at(0:0.5)[phi,label={[label distance=-2pt]-90:$\hspace{10pt}s=3/2$}]{$\Omega$};
                    \node[ellip,fit=(N1)(N2),label={[label distance=-5pt]70:$s=1,t=0$}]{};
                    \node at(180:0.13){\scriptsize$d$};
                \end{scope}
                \node at(0,0.8){$\underbrace{\hspace{3.4cm}}$};
                \node at(-0.1,1.7){$+$};
                \node at(0.2,1.7){$\sqrt{\dfrac{3}{8}}$};
                \node at(-1.8,1.7){$\sqrt{\dfrac{5}{8}}$};
            \end{tikzpicture}&
            \begin{tikzpicture}[N/.style={circle,fill=cyan!40,inner sep=1pt},phi/.style={circle,fill=red!40,inner sep=1pt},ellip/.style={ellipse,draw=gray,line width=0.4pt,dash pattern=on 1pt off 1pt,inner sep=0pt}]
                \node at(0,3){$S=\dfrac{3}{2},T=1$};
                \begin{scope}[shift={(120:2)}]
                    \node(N1)at(120:0.3)[N]{$N$};
                    \node(N2)at(240:0.3)[N,label={[label distance=-2pt]-90:$s=1/2$}]{$N$};
                    \node(phi)at(0:0.3)[phi]{$\Omega$};
                    \node[ellip,rotate fit=60,fit=(N1)(phi),label={[label distance=-3pt]0:$s=1$}]{};
                \end{scope}
                \begin{scope}[shift={(60:2)}]
                    \node(N1)at(120:0.3)[N]{$N$};
                    \node(N2)at(240:0.3)[N,label={[label distance=-2pt]-90:$s=1/2$}]{$N$};
                    \node(phi)at(0:0.3)[phi]{$\Omega$};
                    \node[ellip,rotate fit=60,fit=(N1)(phi),label={[label distance=-3pt]0:$s=2$}]{};
                \end{scope}
                \begin{scope}[shift={(-90:0)}]
                    \node(N1)at(120:0.25)[N]{$N$};
                    \node(N2)at(240:0.25)[N]{$N$};
                    \node(phi)at(0:0.5)[phi,label={[label distance=-2pt]-90:$\hspace{10pt}s=3/2$}]{$\Omega$};
                    \node[ellip,fit=(N1)(N2),label={[label distance=-5pt]70:$s=0,t=1$}]{};
                \end{scope}
                \node at(0,0.8){$\underbrace{\hspace{3.4cm}}$};
                \node at(-0.1,1.7){$+$};
                \node at(-1.9,1.7){$-\sqrt{\dfrac{3}{8}}$};
                \node at(0.2,1.7){$\sqrt{\dfrac{5}{8}}$};
            \end{tikzpicture}
        \end{tabular}
        \caption{Spin-isospin configurations of the $NN\Omega$ system. The upper panels display the $S=1/2$ (left) and $S=5/2$ (right) states. For these two cases, all three coupling schemes shown are equivalent. The lower panels show the $S=3/2$ states, where only the case of the two nucleons coupling first in spin and isospin space is considered, giving rise to $T=0$ (left) and $T=1$ (right) states. Each $S=3/2$ state can then be decomposed into a linear superposition of the other two coupling schemes with the indicated recoupling coefficients. The ellipses indicate spin coupling within two-body subsystems, and $d$ denotes deuteron.}\label{fig:Spin}
	\end{figure}
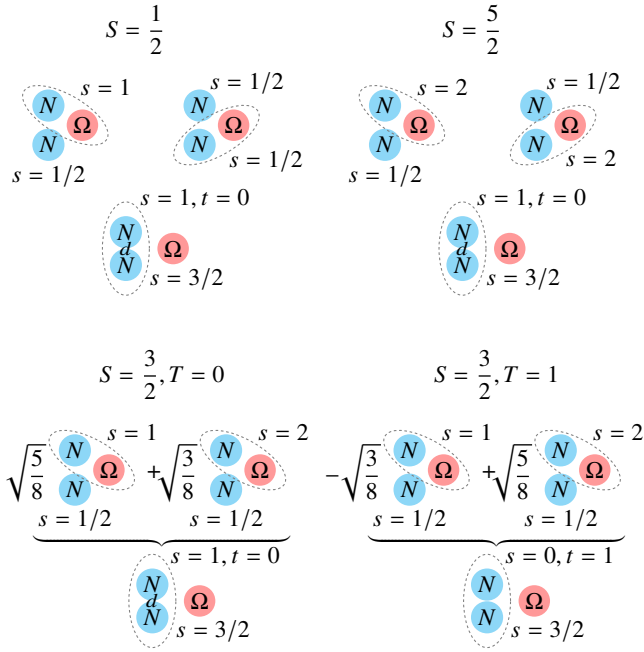

    Since the $\Omega$ baryon carries spin $3/2$, the $NN\Omega$ three-body system admits several distinct spin-isospin states. The two nucleons ($s=1/2$) and the $\Omega$ baryon ($s=3/2$) can couple to total spin $S=1/2$, $S=3/2$, or $S=5/2$, as illustrated in Fig.~\ref{fig:Spin}. For the $S=1/2$ and $S=5/2$ cases, all three coupling schemes shown in the upper panels of Fig.~\ref{fig:Spin} are equivalent. The $S=3/2$ states are more subtle: here we consider only the case where the two nucleons first couple in spin and isospin space, which leads to isosinglet ($T=0$) and isotriplet ($T=1$) states as shown in the lower panels of Fig.~\ref{fig:Spin}. Each $S=3/2$ state can then be decomposed into a linear superposition of the other two coupling schemes with the indicated recoupling coefficients. These coefficients follow directly from the angular-momentum recoupling of the three-spin system and reflect how different $N\Omega$ spin channels contribute to a given $NN\Omega$ state. For the $S=3/2,T=0$ state, 
    \begin{equation}
        V_{N\Omega}(r)=\frac{5}{8}V_{N\Omega}^{J=1}(r)+\frac{3}{8}V_{N\Omega}^{J=2}(r).
    \end{equation}
    For the $S=3/2,T=1$ state, 
    \begin{equation}
        V_{N\Omega}(r)=\frac{3}{8}V_{N\Omega}^{J=1}(r)+\frac{5}{8}V_{N\Omega}^{J=2}(r).
    \end{equation}
    Since the $N\Omega$ interaction is available only in the ${}^{5}S_{2}$ channel, we can compute only this $^{1,6}S_{5/2}$ state for the $NN\Omega$ system.
    
    The $\alpha$ particle is a spin-isospin singlet ($S_\alpha=0$, $T_\alpha=0$), which is precisely the situation encountered in the lower right panels of Fig.~\ref{fig:Spin}: the two nucleons are coupled to spin 0 within the $NN\Omega$ system, just as the four nucleons inside the $\alpha$ cluster are. Therefore, the $\alpha\Omega$ case follows directly from the $S=3/2, T=1$ configurations of the $NN\Omega$ system. In particular, the weights of the $J=1$ and $J=2$ channels in the spin-averaged potential that enters the $\alpha\Omega$ folding procedure can be obtained from the recoupling coefficient in the lower right panels of Fig.~\ref{fig:Spin}, that is
    \begin{equation}\label{eq:Vave}
        V_{N\Omega}^{\text{av}}(r)=\frac{3}{8}V_{N\Omega}^{J=1}(r)+\frac{5}{8}V_{N\Omega}^{J=2}(r).
    \end{equation}
    This is consistent with directly assigning a weight of $2S+1$. For a more detailed discussion, please refer to Ref.~\cite{Zhou:2025anp}. Likewise, the aforementioned discussion is equally applicable to $\Omega_{ccc}$.

{For the $N\Omega$ system, the lattice-QCD input adopted provides the interaction only in the $J=2$ channel. We therefore use this potential as an approximation to the spin-averaged interaction entering the folding procedure. This choice does not imply that the $J=1$ contribution vanishes. Results for related systems offer some guidance regarding the possible effect of this approximation. HAL QCD calculations find stronger attraction in the lower-spin channel for both $N\Omega_{ccc}$ and $NJ/\psi$~\cite{Lyu:2024ttm,Zhang:2025zaa}. A similar trend is suggested for $N\phi$ by the analysis of Ref.~\cite{Chizzali:2022pjd}, which combines the HAL QCD spin-$3/2$ potential with the ALICE correlation data. If this hierarchy also holds for $N\Omega$, including the $J=1$ contribution would strengthen the folded attraction, making the present $J=2$-based calculation a conservative estimate. 
}

\begin{figure}
        \begin{tikzpicture}
			\begin{axis}[
				xlabel=$r$ (fm),
				xmin=0, xmax=5, 
				ylabel=$V(r)$ (MeV),
				ymin=-200, 
				ytick distance=20, minor y tick num=4,
				legend entries={$N\Omega$ HAL QCD potential, $N\Omega$ modified potential~\cite{Filikhin:2025ivu}, $V_{\alpha\Omega}$ ($R_\alpha=1.70$ fm), $\tilde{V}_{\alpha\Omega}$ ($R_\alpha=1.70$ fm), double Gaussian fit~\cite{Filikhin:2025ivu}},
				legend pos=south east,
                legend cell align=left,
				thick
				]
                \newcommand\mpp{139.57039*5.067730716156396*1e-3}
				\newcommand\al{-313.0}
				\newcommand\bl{81.7}
				\newcommand\az{-252}
				\newcommand\bz{0.85}
				\addplot+[domain=0.1:5,samples=300,mark=none,densely dotted,pink]{\al*exp(-\bl*x*x)+\az*(1-exp(-\bz*x*x))*exp(-2*\mpp*x)/x/x};
                \renewcommand\al{-80.289}
				\renewcommand\bl{0.634}
				\renewcommand\az{-60.959}
				\renewcommand\bz{1.131}
				\addplot+[domain=0:5,samples=300,mark=none,dotted,cyan]{\al*exp(-x*x/\bl/\bl)+\az*exp(-x*x/\bz/\bz)};
                \newcommand\Vo{100}
                \newcommand\R{1.33}
                \newcommand\cc{0.583}
				\addplot+[domain=0:5,samples=300,mark=none,red]{-\Vo/(1+exp((x-\R)/\cc))};
				\newcommand\ax{1.70*sqrt(2/3)}
				\addplot+[domain=0:5,samples=300,mark=none,blue,dashed]{4*\al*pow(\bl,3)/pow(\ax*\ax+\bl*\bl,1.5)*exp(-x*x/(\ax*\ax+\bl*\bl))+4*\az*pow(\bz,3)/pow(\ax*\ax+\bz*\bz,1.5)*exp(-x*x/(\ax*\ax+\bz*\bz))};
				\renewcommand\al{-1.288}
				\renewcommand\bl{1.153}
				\renewcommand\az{-23.998}
				\renewcommand\bz{2.624}
				\addplot+[domain=0:5,samples=300,mark=none,black,dash dot]{\al*exp(-x*x/\bl/\bl)+\az*exp(-x*x/\bz/\bz)};
				\draw[gray,thin](axis cs:0,0)--(axis cs:5,0);
			\end{axis}
		\end{tikzpicture}
  \caption{Comparison of the $N\Omega$ and $\alpha\Omega$ potentials from different treatments. The pink dotted curve is the original HAL QCD $N\Omega$ potential in the ${}^5S_2$ channel. The cyan dotted curve is the $N\Omega$ modified potential from Ref.~\cite{Filikhin:2025ivu} with the short-range part removed. The red solid and blue dashed curves are the $\alpha\Omega$ folded potentials obtained from the original HAL QCD and the modified $N\Omega$ potentials, respectively, using the single folding model with $R_\alpha=1.70$~fm. The black dash-dotted curve is the double Gaussian fit to the $r\geq2$~fm tail of the $\alpha\Omega$ potential performed in Ref.~\cite{Filikhin:2025ivu}. We argue that directly using the original HAL QCD potential without any ad~hoc modification is physically more justified.}\label{fig:potential-comparison}
 \end{figure}
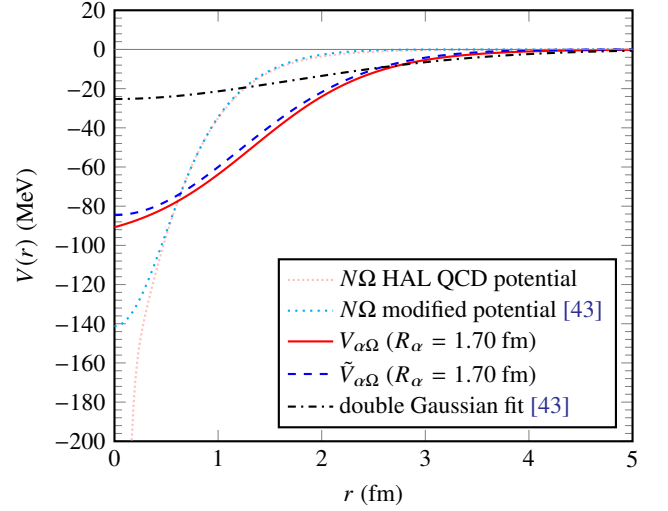

	\subsection{Gaussian expansion method}
	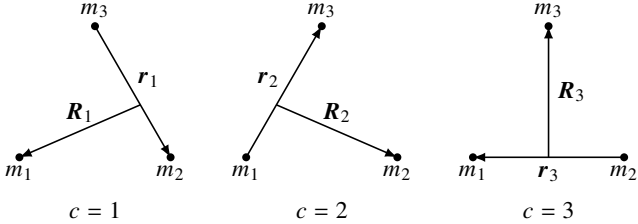
\begin{figure}
		\begin{tikzpicture}[semithick]
			\fill(-1,0)node[below]{$m_1$}circle[radius=1.5pt];
			\fill(1,0)node[below]{$m_2$}circle[radius=1.5pt];
			\fill(0,1.732)node[above]{$m_3$}circle[radius=1.5pt];
			\draw[-latex](-1,0)--node[above left=-1.5pt]{$\boldsymbol{r}_2$}(0,1.732);
			\draw[-latex](-0.6,0.693)--node[above]{$\boldsymbol{R}_2$}(1,0);
			\node at(0,-0.7){$c=2$};
			\begin{scope}[xshift=-3cm]
				\fill(-1,0)node[below]{$m_1$}circle[radius=1.5pt];
				\fill(1,0)node[below]{$m_2$}circle[radius=1.5pt];
				\fill(0,1.732)node[above]{$m_3$}circle[radius=1.5pt];
				\draw[-latex](0,1.732)--node[above right=-1.5pt]{$\boldsymbol{r}_1$}(1,0);
				\draw[-latex](0.6,0.693)--node[above]{$\boldsymbol{R}_1$}(-1,0);
				\node at(0,-0.7){$c=1$};
			\end{scope}
			\begin{scope}[xshift=3cm]
				\fill(-1,0)node[below]{$m_1$}circle[radius=1.5pt];
				\fill(1,0)node[below]{$m_2$}circle[radius=1.5pt];
				\fill(0,1.732)node[above]{$m_3$}circle[radius=1.5pt];
				\draw[-latex](1,0)--node[below]{$\boldsymbol{r}_3$}(-1,0);
				\draw[-latex](0,0)--node[right]{$\boldsymbol{R}_3$}(0,1.732);
				\node at(0,-0.7){$c=3$};
			\end{scope}
		\end{tikzpicture}
		\caption{Three Jacobian coordinates of three-body system.\label{fig:Jacobian coordinates}}
	\end{figure}

For a system described by the Hamiltonian $H$, the stationary Schr\"odinger equation reads
\begin{equation}
H\Psi=E\Psi.
\end{equation}
Within the GEM, the total wave function is expanded in a set of $L^2$-integrable basis functions $\Phi_\alpha$,
\begin{equation}
\Psi=\sum_\alpha C_\alpha\Phi_\alpha,
\end{equation}
where $\alpha$ collectively denotes the quantum numbers specifying each basis state. Applying the Rayleigh-Ritz variational principle leads to the generalized eigenvalue problem
\begin{equation}
HC=NCE,
\end{equation}
with the Hamiltonian and overlap matrix elements defined as
\begin{align}
		&H_{\alpha'\alpha}=\langle\Phi_{\alpha'}|H|\Phi_\alpha\rangle,\label{eq:the energy matrix element}\\
		&N_{\alpha'\alpha}=\langle\Phi_{\alpha'}|\Phi_\alpha\rangle.
	\end{align}
A key feature of the GEM is the use of Gaussian basis functions with ranges chosen over a wide spatial scale, which provides an efficient and flexible representation of few-body wave functions. With an appropriately constructed Gaussian basis, both short-range correlations and long-range spatial behavior can be described accurately, while rapid spatial variations of the wave function can also be efficiently accommodated~\cite{Hiyama:2003cu,Hiyama:2012sma}.

For the $\alpha+\alpha+\Omega/\Omega_{ccc}$ three-body systems, we employ the Jacobi coordinate sets illustrated in Fig.~\ref{fig:Jacobian coordinates}, with the two identical $\alpha$ particles assigned to $m_1$ and $m_2$. Exploiting the exchange symmetry of the two $\alpha$ clusters, the wave function in the third Jacobi coordinate set is expanded in the basis
\begin{equation}\label{eq:Phi1}
	\Phi_\alpha=[\phi_{nl}^{\mathrm{G}}(\boldsymbol{r}_3)\phi_{NL}^{\mathrm{G}}(\boldsymbol{R}_3)]_{L_tM_t},
\end{equation}
where $\alpha={l,L,L_t,n,N}$ specifies the orbital angular momenta and Gaussian basis indices. The angular-momentum configurations and Gaussian basis parameters adopted for each $J^P$ state are summarized in Table~\ref{tab:Angular-momentum space}.

    \begin{table}[htbp]
		\caption{Three-body angular-momentum configuration space of Be. The parameters $\nu_1$ and $\lambda_1$ have units of $\text{fm}^{-2}$; all other quantities are dimensionless. \label{tab:Angular-momentum space}}
		\begin{ruledtabular}
			\begin{tabular}{cccccccccc}
				$J^P$ & $l$ & $L$ & $L_t$ & $n_{\max}$ & $\nu_1$ & $q$ & $N_{\max}$ & $\lambda_1$ & $Q$ \\ \hline
				$0^+$ &  0  &  0  &   0   &     20     &   20    & 0.6 &     20     &     30      & 0.6 \\
				$2^+$ &  2  &  0  &   2   &     20     &   20    & 0.6 &     20     &     30      & 0.6 \\
				$4^+$ &  4  &  0  &   4   &     20     &   20    & 0.6 &     20     &     30      & 0.6
			\end{tabular}
		\end{ruledtabular}
	\end{table}
 
 To accelerate convergence for the $0^+$ state, we employ three Jacobian coordinates. Through coordinate transformations, a $(0,0)$ basis in one coordinate set can incorporate higher partial-wave components such as $(1,1)$, $(2,2)$, etc., in another, thereby improving convergence. The basis functions
 \begin{equation}\label{eq:Phi2}
    \begin{aligned}
        \Phi_\alpha=&C_1\left([\phi_{nl}^{\mathrm{G}}(\boldsymbol{r}_1)\phi_{NL}^{\mathrm{G}}(\boldsymbol{R}_1)]_{L_tM_t}
        +[\phi_{nl}^{\mathrm{G}}(\boldsymbol{r}_2)\phi_{NL}^{\mathrm{G}}(\boldsymbol{R}_2)]_{L_tM_t}\right)\\
        &+C_2[\phi_{nl}^{\mathrm{G}}(\boldsymbol{r}_3)\phi_{NL}^{\mathrm{G}}(\boldsymbol{R}_3)]_{L_tM_t}.
    \end{aligned}
\end{equation}
 For the $2^+$ and $4^+$ states, only the third Jacobi coordinate (Eq.~\eqref{eq:Phi1}) is used, and we restrict the calculation to excitations $l$ in the relative coordinate $\boldsymbol{r}_c$, with the $\Omega$ particle predominantly occupying an $S$ wave~\cite{Lee:2019mlt,Wu:2019ivs}. A detailed discussion of the basis convergence and the contributions of different angular-momentum configurations can be found in Ref.~\cite{Zhou:2025anp}. {For completeness, we have also performed an analogous configuration-space comparison for the $\alpha\alpha\Omega$ system, which leads to the same conclusion concerning the adopted truncation as shown in Table~\ref{tab:Proportion}.

\begin{table*}[htbp]
    \centering
    \caption{Energy levels of $\prescript{9}{\Omega}{\mathrm{Be}}$ obtained using different basis sets with the $V_{\alpha\Omega}^{J=2}(r)$ potential, together with the weight of each basis. Due to the non‑orthogonality of the basis functions defined in different Jacobi coordinates, the last two entries in the second column give the projection of each basis onto the total wave function. The other Gaussian basis parameters are the same as those listed in Table~\ref{tab:Angular-momentum space}. The boldface energy levels correspond to the final values adopted in the article. NBS represents that there are no bound states. \label{tab:Proportion}}
    \begin{tabular}{ccc}
    \hline\hline
    $0^+_1$&$2^+_1$&$4^+_1$\\
    \hline
    \begin{tabular}[t]{crc}
        $(c,l,L)$&Weight&$E$ (MeV)\\
        \hline
        $(3,0,0)$&100.0&$-$44.0\\[10pt]
        $(3,0,0)$&100.0&\multirow{1.5}{*}{$-$44.0}\\
        $(3,1,1)$&0.00&\\[10pt]
        $(3,0,0)$&94.74&\\
        $(3,1,1)$&0.00&$-$46.3\\
        $(3,2,2)$&5.26&\\[10pt]
        $(1,0,0)$&\multirow{1.5}{*}{21.15}&\\
        $(2,0,0)$&&$\boldsymbol{-46.4}$\\
        $(3,0,0)$&78.85&\\
    \end{tabular}&
    \begin{tabular}[t]{crc}
        $(c,l,L)$&Weight&$E$ (MeV)\\
        \hline
        $(3,2,0)$&100.0&$\boldsymbol{-44.3}$\\[10pt]
        $(3,0,2)$&100.0&$-$2.10\\[10pt]
        $(3,2,0)$&99.31&\multirow{1.5}{*}{$-$44.6}\\
        $(3,0,2)$&0.69&\\[10pt]
        $(1,2,0)$&\multirow{1.5}{*}{0.47}&\\
        $(2,2,0)$&&$-$44.3\\
        $(3,2,0)$&99.53&\\
    \end{tabular}&
    \begin{tabular}[t]{crc}
        $(c,l,L)$&Weight&$E$ (MeV)\\
        \hline
        $(3,4,0)$&100.0&$\boldsymbol{-46.6}$\\[10pt]
        $(3,2,2)$&100.0&$-$1.61\\[10pt]
        $(3,0,4)$&100.0&NBS\\[10pt]
        $(3,4,0)$&99.50&\\
        $(3,2,2)$& 0.50&$-46.9$\\
        $(3,0,4)$& 0.00&\\
    \end{tabular}\\
    \hline\hline
    \end{tabular}
\end{table*}
 }

 The Gaussian basis functions appearing in Eq.~\eqref{eq:Phi2} are written as
	\begin{equation}
		\begin{array}{ll}
			\phi_{nlm}^{\mathrm{G}}(\boldsymbol{r})=\phi_{nl}^{\mathrm{G}}(r)Y_{lm}(\hat{\boldsymbol{r}}),&\phi_{nl}^{\mathrm{G}}(r)=N_{nl}r^l\mathrm{e}^{-\nu_nr^2},\\
			\phi_{NLM}^{\mathrm{G}}(\boldsymbol{R})=\phi_{NL}^{\mathrm{G}}(R)Y_{LM}(\hat{\boldsymbol{R}}),&\phi_{NL}^{\mathrm{G}}(R)=N_{NL}R^L\mathrm{e}^{-\lambda_NR^2},
		\end{array}
	\end{equation}
	where the normalization constants are
	\begin{equation}
		N_{nl}=\left(\frac{2^{l+2}(2\nu_n)^{l+\frac{3}{2}}}{\sqrt{\pi}(2l+1)!!}\right)^{\frac{1}{2}},\quad N_{NL}=\left(\frac{2^{L+2}(2\lambda_N)^{L+\frac{3}{2}}}{\sqrt{\pi}(2L+1)!!}\right)^{\frac{1}{2}}.
	\end{equation}
	 The Gaussian range parameters are taken in geometric progressions
	\begin{equation}
		\begin{array}{ll}
			\nu_n=\nu_1q^{n-1}&(n=1\sim n_{\max}),\\
			\lambda_N=\lambda_1Q^{N-1}&(N=1\sim N_{\max}).
		\end{array}
	\end{equation}
 Such geometric progressions allow the Gaussian basis to efficiently cover a wide range of length scales and provide rapid numerical convergence within the GEM~\cite{Hiyama:2003cu,Hiyama:2012sma}.
	
	\subsection{Complex scaling method}
The CSM provides a convenient framework for treating resonant states by applying a complex rotation to the Jacobi coordinates,
\begin{equation}
r_c\to r_c\mathrm{e}^{\mathrm{i}\theta},
\qquad
R_c\to R_c\mathrm{e}^{\mathrm{i}\theta},
\end{equation}
where $\theta$ is the scaling angle. Under this transformation, the kinetic and potential terms of the Hamiltonian are transformed as
\begin{equation}
T\to T\mathrm{e}^{-2\mathrm{i}\theta},
\qquad
V_{ij}(r_{ij})\to
V_{ij}(r_{ij}\mathrm{e}^{\mathrm{i}\theta}).
\end{equation}

Solving the resulting complex-scaled Schr\"odinger equation yields a set of complex energy eigenvalues. Bound states remain on the negative real axis, while continuum eigenvalues rotate into the lower half of the complex-energy plane by an angle $2\theta$. In contrast, resonance poles remain approximately stationary with respect to variations of $\theta$ once they are isolated from the rotated continua. The complex energy of a resonance is written as
\begin{equation}
E=E_r-\mathrm{i}\Gamma_r/2,
\end{equation}
where $E_r$ and $\Gamma_r$ denote the resonance energy and decay width, respectively. The complex-scaled Schr\"odinger equation is solved within the same GEM framework used for the bound-state calculations.

To characterize the spatial structure of each state, we evaluate the corresponding rms interparticle distance. For a bound state, it is defined as
\begin{equation}
R=\sqrt{\langle\Psi|r_{ij}^2|\Psi\rangle}.
\end{equation}
For a resonant state in the CSM, we use
\begin{equation}
R=
\operatorname{Re}\left[
\sqrt{
\frac{
(\Psi(\theta)|r_{ij}^2\mathrm{e}^{2\mathrm{i}\theta}|\Psi(\theta))
}{
(\Psi(\theta)|\Psi(\theta))
}}
\right],
\end{equation}
where $\Psi(\theta)$ denotes the complex-scaled wave function obtained at the angle $\theta$. The round inner product is defined as~\cite{Romo:1968tcz}
\begin{equation}
(\psi|\phi)
=
\int
\psi(\boldsymbol{r})
\phi(\boldsymbol{r})
,\mathrm{d}\boldsymbol{r},
\end{equation}
without complex conjugation of the bra vector. The rms radius obtained in this way is generally complex for a resonant state, whereas it is real for a bound state. In practice, the extracted rms radius approaches a stable value as the scaling angle is increased, consistent with the behavior discussed in Ref.~\cite{Homma:1997wtc}.

	\section{numerical results}\label{sec3}
 
To obtain more stable results, bound states were calculated by solving the Schr\"odinger equation using the GEM, while resonant states were obtained by solving the complex-scaled Schr\"odinger equation with the same method.

The energy spectra of $\prescript{8}{}{\text{Be}}$, $\prescript{9}{\Lambda}{\text{Be}}$, $\prescript{8}{\phi}{\text{Be}}$, $\prescript{9}{\Omega}{\text{Be}}$, $\prescript{9}{J/\psi}{\text{Be}}$, and $\prescript{9}{\Omega_{ccc}}{\text{Be}}$ are shown in Fig.~\ref{fig:energy spectra}. The energy spectrum of $\prescript{8}{}{\text{Be}}$ was computed using a phenomenological $\alpha\alpha$ potential with the CSM and GEM and agrees well with the experimental values of 0.0918, 3.03, and 11.35 MeV for the $0^+_1$, $2^+_1$, and $4^+_1$ states, respectively~\cite{Tilley:2004zz}. The energy spectrum of $\prescript{9}{\Lambda}{\text{Be}}$ is taken from Refs.~\cite{Lee:2019mlt,Wu:2019ivs}, where it was derived using the OCM and CSM. The energy spectra of $\prescript{9}{\Omega}{\text{Be}}$ and $\prescript{9}{\Omega_{ccc}}{\text{Be}}$ were calculated via GEM and CSM using the HAL QCD potentials and a phenomenological $\alpha\alpha$ potential, with the $\alpha$-particle root-mean-square matter radius $R_\alpha$ taken as 1.84, 1.70, and 1.56 fm, respectively. For comparison, we also include the results for $\prescript{8}{\phi}{\text{Be}}$ and $\prescript{9}{J/\psi}{\text{Be}}$ obtained using the same methodology from Ref.~\cite{Zhou:2025anp}. Figure~\ref{fig:energy spectra} displays the case for $R_\alpha = 1.70$ fm.

Table~\ref{tab:energy spectra} summarizes the energies and rms interparticle distances of the $N\Omega$, $NN\Omega$, $\alpha+\Omega$ ($\prescript{5}{\Omega}{\mathrm{He}}$), $\alpha+\alpha$ ($\prescript{8}{}{\mathrm{Be}}$), and $\alpha+\alpha+\Omega$ ($\prescript{9}{\Omega}{\mathrm{Be}}$) systems and their $\Omega_{ccc}$ analogues. To characterize the structural response induced by the $\Omega$ and $\Omega_{ccc}$ baryons, we also evaluate the relevant rms interparticle distances, in particular $R_{\alpha\Omega}$ and $R_{\alpha\alpha}$. For quantities that depend on the $\alpha$-particle rms matter radius $R_\alpha$, the values at the reference choice $R_\alpha=1.70$~fm are quoted explicitly, while the superscripts and subscripts correspond to the results obtained with $R_\alpha=1.84$ and $1.56$~fm, respectively. The systematic dependence of the energies and spatial distributions on $R_\alpha$ and orbital angular momentum follows the same general behavior discussed in our previous study~\cite{Zhou:2025anp}; here we therefore focus on the distinctive dynamical effects associated with the $\Omega$ and $\Omega_{ccc}$ baryons.

	\begin{figure*}[htbp]
        \begin{tikzpicture}
			\begin{axis}[
				width=18.6cm, height=11cm,
				axis line style={thick},
				ylabel=$E$ (MeV),
                ylabel style={at={(axis cs:0,16)},rotate=-90},
				ymin=-14.99, ymax=15,
				extra y ticks={-11},       
				extra y tick labels={$-$44},
				ytick distance=5, minor y tick num=4,
				ytick style={thick, black, line cap=round},
				yticklabel style={/pgf/number format/.cd, fixed, precision=0, fixed zerofill},
				major tick length=0.13cm,
				minor tick length=0.07cm,
				xmin=-0.5, xmax=8.6,
				xtick={0,1,2,3,4,5,6,7,8},
				xticklabels={$\prescript{8}{}{\text{Be}}$,$\prescript{9}{\Lambda}{\text{Be}}$~\cite{Lee:2019mlt,Wu:2019ivs},$\prescript{8}{\phi}{\text{Be}}(V_{\alpha\phi}^{3/2})$,$\prescript{9}{\Omega}{\text{Be}}(V_{\alpha\Omega}^{J=2})$,$\prescript{9}{J/\psi}{\text{Be}}(V_{\alpha J/\psi}^{\text{av}})$,$\prescript{9}{\Omega_{ccc}}{\text{Be}}(V^{J=2})$,$\prescript{9}{\Omega_{ccc}}{\text{Be}}(V^{\text{av}})$,$\prescript{9}{\Omega_{ccc}}{\text{Be}}(V^{J=1})$,continuum},
				xtick style={draw=none},
				nodes near coords,
				nodes near coords align={above=-1.5pt},
				coordinate style/.condition={y==-0.156||y==-5.34||y==-0.826||y==1.94||y==-1.05||y==2.11||y==-1.54||y==-0.219||y==2.59||y==11.0||y==-46.6+33}{below=-1.5pt},
				coordinate style/.condition={y==-21.1+4}{left=14},
				]
				\addplot[mark=-, mark size=0.5cm, semithick, line cap=round,
				draw=none, point meta=explicit symbolic,
				]coordinates
				{
					(0,0.09)[0.09 (1E-4)]
					(0,2.90)[2.90 (1.3)]
					(0,11.6)[11.6 (3.1)]
					(1,-6.65)[$-$6.65]
					(1,-3.82)[$-$3.82]
					(1,3.2)[3.2 (0.78)]
					(2,-9.34)[$-$9.34]
					(2,-5.34)[$-$5.34]
					(2,-0.156)[$-$0.156]
                    (3,-46.4+33)[$-$46.4]
					(3,-44.3+33)[$-$44.3]
					(3,-46.6+33)[$-$46.6]
					(4,-1.05)[$-$1.05]
					(4,1.94)[1.94 (0.33)]
					(4,9.86)[9.86 (1.7)]
					(5,-1.84)[$-$1.84]
					(5,1.49)[1.49 (0.10)]
					(5,8.87)[8.87 (1.1)]
                    (6,-2.40)[$-$2.40]
					(6,1.08)[1.08 (0.028)]
					(6,8.20)[8.20 (0.85)]
					(7,-3.43)[$-$3.43]
					(7,0.254)[0.254 (4.7E-3)]
					(7,7.00)[7.00 (0.46)]
				};
				\node at(axis cs:0.55,0.8){$0^+_1$};
				\node at(axis cs:0.5,3.4){$2^+_1$};
				\node at(axis cs:0.5,12.2){$4^+_1$};
				\node at(axis cs:1.42,-6.2){$0^+_1$};
				\node at(axis cs:1.42,-3.4){$2^+_1$};
				\node at(axis cs:1.50,3.7){$4^+_1$};
				\node at(axis cs:2.4,-9){$0^+_1$};
				\node at(axis cs:2.4,-6.2){$2^+_1$};
				\node at(axis cs:2.4,-1){$4^+_1$};
				\node at(axis cs:3.6,-1.8){$0^+_1$};
				\node at(axis cs:3.52,1.2){$2^+_1$};
				\node at(axis cs:3.55,10.5){$4^+_1$};
				\node at(axis cs:3.4,-46+33){$0^+_1$};
				\node at(axis cs:3.4,-43.8+33){$2^+_1$};
				\node at(axis cs:3.4,-47.3+33){$4^+_1$};
				\node at(axis cs:4.6,-1.5){$0^+_1$};
				\node at(axis cs:4.55,2.0){$2^+_1$};
				\node at(axis cs:4.55,9.5){$4^+_1$};
				\node at(axis cs:5.6,-2.00){$0^+_1$};
				\node at(axis cs:5.5,1.6){$2^+_1$};
				\node at(axis cs:5.5,8.7){$4^+_1$};
				\node at(axis cs:6.6,-3){$0^+_1$};
				\node at(axis cs:6.4,0.5){$2^+_1$};
				\node at(axis cs:6.52,7.5){$4^+_1$};
				%
				%
				\draw[dashed] (axis cs:-0.5,0)--(axis cs:7.6,0)node[right]{\footnotesize$\alpha+\alpha+\Omega$};
				\draw[dashed] (axis cs:0.7,2.9)--(axis cs:7.6,2.9)node[right]{\footnotesize$\prescript{8}{}{\text{Be}}(2^+)+\Omega$};
				\draw[dashed] (axis cs:0.7,11.6)--(axis cs:7.6,11.6)node[right]{\footnotesize$\prescript{8}{}{\text{Be}}(4^+)+\Omega$};
			\end{axis}
		\end{tikzpicture}
		\caption{Energy spectra of $\prescript{8}{}{\text{Be}}$ obtained with the phenomenological $\alpha\alpha$ potential, $\prescript{9}{\Lambda}{\text{Be}}$ from Refs.~\cite{Lee:2019mlt,Wu:2019ivs}, and $\prescript{8}{\phi}{\text{Be}}$ and $\prescript{8}{J/\psi}{\text{Be}}$ from Ref.~\cite{Zhou:2025anp}, together with those of $\prescript{9}{\Omega}{\text{Be}}$ and $\prescript{9}{\Omega_{ccc}}{\text{Be}}$ calculated in the present work for $R_{\alpha}=1.70$ fm. Values in parentheses accompanying the energy levels denote the decay widths, and the dashed horizontal lines indicate the corresponding decay thresholds. Parenthetical labels following the nuclei specify the $\alpha$-hadron interactions employed, while for $\prescript{9}{\Omega_{ccc}}{\mathrm{Be}}$ the subscript $\alpha\Omega_{ccc}$ is omitted for brevity. The interval between $-44$ and $-10$ MeV on the vertical axis is compressed for clarity. \label{fig:energy spectra}}
	\end{figure*}

 	\begin{table*}[htbp]
		\caption{Energies and rms interparticle distances of the $N\Omega$, $NN\Omega$, $\alpha+\Omega$ ($\prescript{5}{\Omega}{\text{He}}$), $\alpha+\alpha$ ($^{8}\mathrm{Be}$), and $\alpha+\alpha+\Omega$ ($\prescript{9}{\Omega}{\text{Be}}$) systems and their $\Omega_{ccc}$ analogues. For resonant states, the decay widths are given in parentheses following the resonance energies. The quantities $R_{N\Omega}$, $R_{NN}$, $R_{\alpha\Omega}$, and $R_{\alpha\alpha}$ denote the corresponding rms interparticle distances. For entries that depend on the $\alpha$-particle rms matter radius $R_\alpha$, the values at the reference choice $R_\alpha=1.70$ fm are quoted explicitly, with the superscripts and subscripts corresponding to the results obtained for $R_\alpha=1.84$ and $1.56$ fm, respectively. NBS indicates that no bound state is found. \label{tab:energy spectra}}
\begin{tabular*}{\textwidth}{@{\extracolsep{\fill}}cccccccc}
\hline\hline
		                                                                     \multicolumn{6}{c}{$N\Omega$ and $N\Omega_{ccc}$}                                                                      &      \multicolumn{2}{c}{$NN$}       \\ \cline{1-6}\cline{7-8}
	Systems              & $^{2S+1}L_J$  & $E$ (MeV) & Refs.~\cite{HALQCD:2018qyu,Zhang:2025zaa} & $R_{N\Omega}$ (fm) & Refs.~\cite{HALQCD:2018qyu,Zhang:2025zaa} & $R_{NN}$ (fm) & $E$ (MeV) \\ \hline
	$n\Omega$             &    $^5S_2$    &  $-$2.19  &                 $-$1.54~\cite{HALQCD:2018qyu}                  &          3.29           &                   3.77~\cite{HALQCD:2018qyu}                   &                  3.98    &    $-$2.23   \\
	$p\Omega^-$            &    $^5S_2$    &  $-$3.16  &                 $-$3.00~\cite{HALQCD:2018qyu}                  &          2.94           &                   3.01~\cite{HALQCD:2018qyu}                   &                         &           \\
	$n\Omega_{ccc}/p\Omega_{ccc}^{++}$ & $^5S_2/^3S_1$ &    NBS    &                   NBS~\cite{Zhang:2025zaa}                    &           NBS           &                   NBS~\cite{Zhang:2025zaa}                    &                         &\\                                                                                                                                                                                                                \multicolumn{8}{c}{$NN\Omega$ and $NN\Omega_{ccc}$}                                                                                                                                                                                                                               \\ \hline
	    Systems     &                                 $^{2T+1,2S+1}L_J$                                  & $E$ (MeV) &                                     Refs.~\cite{Garcilazo:2018gkb,Garcilazo:2019igo,Zhang:2021vsf,Etminan:2023gzh,Filikhin:2025nvt,Etminan:2025emv,Filikhin:2025ige,Filikhin:2025ivu}                                      & $R_{N\Omega}$ (fm) &       Refs.~\cite{Etminan:2023gzh,Filikhin:2025nvt}        & $R_{NN}$ (fm) &       Refs.~\cite{Etminan:2023gzh,Filikhin:2025nvt}        \\ \hline
	  $NN\Omega$    &                                  $^{1,6}S_{5/2}$                                   &  $-$23.5  & \makecell[c]{$-$16.34~\cite{Garcilazo:2018gkb}, $-$21.3~\cite{Garcilazo:2019igo},\\ $-$22.0~\cite{Zhang:2021vsf}, $-$20.953~\cite{Etminan:2023gzh},\\ $-$20.010~\cite{Filikhin:2025nvt}, $-$19.58~\cite{Filikhin:2025ivu}} &          1.69           & 1.768~\cite{Etminan:2023gzh}, 1.77~\cite{Filikhin:2025nvt} &     1.95      & 2.001~\cite{Etminan:2023gzh}, 2.02~\cite{Filikhin:2025nvt} \\
	$NN\Omega_{ccc}$ & \makecell[c]{$^{1,6}S_{5/2}$, $^{1,4}S_{3/2}$,\\ $^{3,4}S_{3/2}$, $^{1,2}S_{1/2}$} &    NBS    &                                                                                        NBS~\cite{Etminan:2025emv,Filikhin:2025ige}                                                                                         &           NBS           &                                                            &      NBS      &\\ \hline
\end{tabular*}
\begin{tabular*}{\textwidth}{@{\extracolsep{\fill}}ccccccc}
	                                        &         &                \multicolumn{3}{c}{$\alpha+\Omega$ and $\alpha+\Omega_{ccc}$ ($\prescript{5}{\Omega}{\text{He}}$ and $\prescript{5}{\Omega_{ccc}}{\text{He}}$)}                & \multicolumn{2}{c}{$\alpha+\alpha\ (\prescript{8}{}{\text{Be}})$} \\ \cline{3-5}\cline{6-7}
	              Interaction               &         &                  $E$ (MeV)                   & Refs.~\cite{Etminan:2019gds,Filikhin:2024hrd,Etminan:2024uvc,Filikhin:2025ivu} &   $R_{\alpha\Omega}$ (fm)   &        $E$ (MeV)        &         $R_{\alpha\alpha}$ (fm)         \\ \hline
	      $V_{\alpha\Omega}^{J=2}(r)$       &         &            $-33.2^{+5.6}_{-7.0}$             &           \makecell[c]{$-$22.8~\cite{Etminan:2019gds}, $-$28.096~\cite{Filikhin:2024hrd}, \\$-$22.9~\cite{Etminan:2024uvc}, $-$6.16~\cite{Filikhin:2025ivu}}            & $1.35^{+0.10}_{-0.10}$  & 0.09 ($1\times10^{-4}$) &                  5.96                   \\[10pt]
	   $V_{\alpha\Omega_{ccc}}^{J=2}(r)$    &         &          $-0.571^{+0.240}_{-0.380}$          &           $-$3.53~\cite{Filikhin:2025ivu}            & $3.67^{+0.95}_{-0.70}$  &       2.90 (1.3)        &                  3.60                   \\
	$V_{\alpha\Omega_{ccc}}^{\text{av}}(r)$ &         &          $-0.939^{+0.349}_{-0.532}$          &                              & $3.05^{+0.64}_{-0.50}$  &       11.6 (3.1)        &                  2.91                   \\
	   $V_{\alpha\Omega_{ccc}}^{J=1}(r)$    &         &           $-1.70^{+0.55}_{-0.81}$            &           $-$9.90~\cite{Filikhin:2025ivu}            & $2.48^{+0.42}_{-0.35}$  &                         &                                         \\ \hline
	                                        &                                                   \multicolumn{6}{c}{$\alpha+\alpha+\Omega$ and $\alpha+\alpha+\Omega_{ccc}$ ($\prescript{9}{\Omega}{\text{Be}}$ and $\prescript{9}{\Omega_{ccc}}{\text{Be}}$)}                                                    \\ \cline{2-7}
	              Interaction               &  State  &                  $E$ (MeV)                   & Ref.~\cite{Filikhin:2025ivu} &   $R_{\alpha\Omega}$ (fm)   &                         &         $R_{\alpha\alpha}$ (fm)         \\ \hline
	      $V_{\alpha\Omega}^{J=2}(r)$       & $0^+_1$ &            $-46.4^{+5.9}_{-7.0}$             &           $-$12.9            & $2.07^{+0.06}_{-0.05}$  &                         &         $2.94^{+0.07}_{-0.07}$          \\
	                                        & $2^+_1$ &            $-44.3^{+6.1}_{-7.2}$             &                              & $1.83^{+0.09}_{-0.09}$  &                         &         $2.64^{+0.09}_{-0.10}$          \\
	                                        & $4^+_1$ &            $-46.6^{+8.1}_{-9.7}$             &                              & $1.61^{+0.09}_{-0.08}$  &                         &         $2.20^{+0.08}_{-0.09}$          \\[10pt]
	   $V_{\alpha\Omega_{ccc}}^{J=2}(r)$    & $0^+_1$ &           $-1.84^{+0.37}_{-0.49}$            &           $-$7.39            & $3.38^{+0.16}_{-0.14}$  &                         &         $4.16^{+0.07}_{-0.06}$          \\
	                                        & $2^+_1$ & $1.49^{+0.20}_{-0.23}(0.10^{+0.07}_{-0.05})$ &                              & $3.54^{+0.05}_{-0.09}$ &                         &        $4.29^{-0.12}_{-0.01}$         \\
	                                        & $4^+_1$ & $8.87^{+0.41}_{-0.49}(1.1^{+0.3}_{-0.3})$ &                              & $2.59^{+0.09}_{-0.08}$ &                         &        $3.00^{-0.05}_{+0.03}$        \\[10pt]
	$V_{\alpha\Omega_{ccc}}^{\text{av}}(r)$ & $0^+_1$ &           $-2.40^{+0.47}_{-0.62}$            &                              & $3.23^{+0.13}_{-0.12}$  &                         &         $4.08^{+0.06}_{-0.06}$          \\
	                                        & $2^+_1$ & $1.08^{+0.26}_{-0.32}(0.028^{+0.041}_{-0.021})$ &                              & $3.42^{+0.14}_{-0.17}$ &                         &         $4.29^{+0.09}_{-0.11}$         \\
	                                        & $4^+_1$ & $8.20^{+0.50}_{-0.61}(0.85^{+0.30}_{-0.28})$ &                              & $2.53^{+0.08}_{-0.08}$ &                         &        $3.06^{-0.04}_{+0.00}$        \\[10pt]
	   $V_{\alpha\Omega_{ccc}}^{J=1}(r)$    & $0^+_1$ &           $-3.43^{+0.64}_{-0.85}$            &           $-$18.0            & $3.05^{+0.11}_{-0.09}$  &                         &         $3.97^{+0.06}_{-0.05}$          \\
	                                        & $2^+_1$ & $0.254^{+0.399}_{-0.446}(0.0047^{+0.0063}_{-0.0047})$ &                              & $3.03^{+0.28}_{-0.19}$ &                         &         $3.93^{+0.31}_{-0.19}$         \\
	                                        & $4^+_1$ & $7.00^{+0.68}_{-0.84}(0.46^{+0.28}_{-0.23})$ &                              & $2.43^{+0.08}_{-0.10}$ &                         &         $3.09^{+0.01}_{-0.04}$\\[3pt]
                                         \hline\hline
\end{tabular*}
	\end{table*}

 \subsection{The $NN\Omega$ and $NN\Omega_{ccc}$ systems}

	Before discussing the $\alpha\alpha\Omega$ and $\alpha\alpha\Omega_{ccc}$ three-body systems, it is instructive to first examine the simpler two-body $N\Omega/\Omega_{ccc}$ and three-body $NN\Omega/\Omega_{ccc}$ systems, whose energy spectra are also listed in Table~\ref{tab:energy spectra}. These results provide a baseline for understanding the strength of the $N\Omega$ and $N\Omega_{ccc}$ interactions and serve as a consistency check against previous studies.

For the $N\Omega$ system in the ${}^{5}S_{2}$ channel, we obtain a shallow bound state with $E=-2.19$ MeV for $n\Omega$, compared with $-1.54$ MeV reported in Ref.~\cite{HALQCD:2018qyu}. The corresponding rms distances are $3.29$ and $3.77$ fm, respectively. For the $p\Omega^{-}$ system, including the Coulomb attraction gives a binding energy of $-3.16$ MeV and an rms distance of $2.94$ fm, which are close to the values $-3.00$ MeV and $3.01$ fm obtained in Ref.~\cite{HALQCD:2018qyu}. The small quantitative differences between our results and those of Ref.~\cite{HALQCD:2018qyu} mainly originate from the different masses used in solving the Schr\"odinger equation. In the present calculation, we employ the physical hadron masses, whereas Ref.~\cite{HALQCD:2018qyu} uses the masses corresponding to the lattice QCD simulation. Nevertheless, the two calculations give a consistent picture of a shallow and spatially extended $N\Omega$ bound state. 

A much stronger binding effect emerges when a second nucleon is added. For the $NN\Omega$ system in the ${}^{1,6}S_{5/2}$ channel, we obtain $E=-23.5$ MeV, which may be compared with the previous predictions ranging from $-16.34$ to $-22.0$ MeV~\cite{Garcilazo:2018gkb,Garcilazo:2019igo,Zhang:2021vsf,Etminan:2023gzh,Filikhin:2025nvt,Filikhin:2025ivu}. Our result therefore lies on the more deeply bound side, but remains at the same characteristic energy scale as the existing few-body calculations. More notably, the calculated rms distances, $R_{N\Omega}=1.69$ fm and $R_{NN}=1.95$ fm, are very close to the values $R_{N\Omega}\simeq1.77$ fm and $R_{NN}\simeq2.00$ fm reported in Refs.~\cite{Etminan:2023gzh,Filikhin:2025nvt}. The simultaneous increase in binding and reduction of the interparticle distances from $N\Omega$ to $NN\Omega$ demonstrate that the attractive $N\Omega$ interaction can generate a compact few-baryon configuration once more than one nucleon participates.

This observation is also relevant to the $\alpha\Omega$ system discussed below. Since both the $N\Omega$ and $NN\Omega$ systems already exhibit appreciable attraction, with the latter reaching a binding scale of several tens of MeV, a relatively deeply bound $\prescript{5}{\Omega}{\mathrm{He}}$ state obtained by embedding the $\Omega$ baryon in an $\alpha$ cluster is qualitatively consistent with the behavior observed in these simpler few-baryon systems.

The situation is qualitatively different for the $\Omega_{ccc}$ sector. Neither $n\Omega_{ccc}$ nor $p\Omega_{ccc}^{++}$ supports a bound state in the ${}^{5}S_{2}$ or ${}^{3}S_{1}$ channel. Moreover, no bound state is found for $NN\Omega_{ccc}$ in any of the four spin-isospin configurations considered, ${}^{1,6}S_{5/2}$, ${}^{1,4}S_{3/2}$, ${}^{3,4}S_{3/2}$, and ${}^{1,2}S_{1/2}$, in agreement with Refs.~\cite{Etminan:2025emv,Filikhin:2025ige}. The absence of binding at both the two- and three-body levels provides a clear contrast with the $\Omega$ sector and indicates that the $N\Omega_{ccc}$ attraction is not strong enough to generate a bound few-baryon system.
    
	\subsection{The $\prescript{5}{\Omega}{\mathrm{He}}$ and $\prescript{9}{\Omega}{\text{Be}}$ systems}
As shown in Fig.~\ref{fig:energy spectra}, the $\Omega$ baryon exhibits an exceptionally strong gluelike effect. Using the $V_{\alpha\Omega}^{J=2}(r)$ potential, the $0^+_1$, $2^+_1$, and $4^+_1$ states of $\prescript{9}{\Omega}{\text{Be}}$ are all deeply bound, with binding energies of approximately $-46.4$, $-44.3$, and $-46.6$~MeV, respectively. These binding energies are nearly an order of magnitude larger than those of $\prescript{9}{\Lambda}{\text{Be}}$ and $\prescript{8}{\phi}{\text{Be}}$, reflecting the extremely strong $N\Omega$ attraction in the ${}^5S_2$ channel. The three states are almost degenerate in energy, indicating that the $\alpha\Omega$ interaction dominates the dynamics of $\prescript{9}{\Omega}{\text{Be}}$, while the $\alpha\alpha$ interaction plays a secondary role. Interestingly, the $4^+_1$ state is predicted to lie slightly below the $0^+_1$ state, resulting in an unconventional level inversion. 


{The evolution of the spectrum with the $\alpha\Omega$ interaction strength clarifies the unconventional level ordering. Scaling the $V_{\alpha\Omega}^{J=2}$ potential by a factor $\lambda$, while keeping the $\alpha\alpha$ interaction unchanged, produces successive level crossings. Figure~\ref{fig:EnergyVary} shows a progressive reordering of the entire spectrum. As $\lambda$ increases, the $4^+$ level first passes below the $2^+$ level and then below the $0^+$ level; subsequently, the $2^+$ level also passes below the $0^+$ level. The slight $0^+$--$4^+$ inversion at the original interaction strength is an intermediate stage of this broader evolution toward a complete reversal of the initial ordering. Table~\ref{tab:EnergyVary} shows that, as $\lambda$ increases from 0.6 to 1.5, the total kinetic energies ($T_{\mathrm{eff}}$ plus the centrifugal contribution) increase by approximately 18.2, 25.4, and 43.9 MeV for the $0^+$, $2^+$, and $4^+$ states, respectively, while their total potential energies decrease by 83.7, 95.1, and 125.6 MeV. The larger potential-energy gains therefore outweigh the larger kinetic-energy costs of the higher partial waves. 

The $\alpha\Omega$ contribution becomes more attractive by approximately 90.0, 98.8, and 109.0 MeV for the three states. Meanwhile, the $\alpha\alpha$ contribution becomes less favorable by 6.3 MeV for the $0^+$ state and by 3.7 MeV for the $2^+$ state, but more favorable by 16.6 MeV for the $4^+$ state. These contributions explain both the rapid lowering of the $4^+$ level and the subsequent crossing between the $2^+$ and $0^+$ levels. This behavior is associated with the partial-wave dependence of the adopted Ali-Bodmer interaction and the structural contraction induced by the $\alpha\Omega$ attraction. The strengths of the short-range nuclear repulsive term are 500, 320, and 0 MeV for $l=0$, 2, and 4, respectively. Increasing the attraction therefore induces different spatial and energetic responses in the three states.

\begin{figure}
\begin{tikzpicture}
  \begin{groupplot}[
    group style={
      group size=1 by 3,
      vertical sep=4mm
    },
    width=7.5cm,
    scale only axis,
    xmin=-0.6,xmax=4,
    enlargelimits=false,
    axis x line=none,
    axis y line*=box,
    axis line style={thick},
    ytick pos=left,
    ytick distance=5,
    minor y tick num=4,
    ytick style={thick,black,line cap=round},
    tick label style={font=\small},
    yticklabel={$\pgfmathprintnumber[fixed,precision=0]{\tick}$},
    scaled y ticks=false,
    major tick length=0.13cm,
    minor tick length=0.07cm,
    clip=false,
    nodes near coords,
    nodes near coords align={above=-1.5pt},
    coordinate style/.condition={y==-39.4||y==-46.6||y==-86.3}{below=-1.5pt},
    coordinate style/.condition={y==-37.8}{left=14},
  ]

    \nextgroupplot[hide x axis=false,
      axis x line*=top,xtick style={draw=none},xticklabels={},
      ymin=-22,ymax=-10,height=2.75cm]
      \addplot[mark=-, mark size=0.5cm, semithick, line cap=round,
				draw=none, point meta=explicit symbolic,
				]coordinates
				{
                    (0,-20.1)[$-$20.1]
					(0,-16.5)[$-$16.5]
					(0,-13.4)[$-$13.4]
                    
				};
				\node at(axis cs:0.5,-19.3){$0^+_1$};
				\node at(axis cs:0.5,-15.7){$2^+_1$};
				\node at(axis cs:0.5,-12.7){$4^+_1$};
    \nextgroupplot[ymin=-49,ymax=-34,height=3.75cm]
      \addplot[mark=-, mark size=0.5cm, semithick, line cap=round,
				draw=none, point meta=explicit symbolic,
				]coordinates
				{
                    (1,-39.4)[$-$39.4]
					(1,-36.7)[$-$36.7]
					(1,-37.8)[$-$37.8]
                    (2,-46.4)[$-$46.4]
					(2,-44.3)[$-$44.3]
					(2,-46.6)[$-$46.6]
				};
				\node at(axis cs:1.5,-40.6){$0^+_1$};
				\node at(axis cs:1.5,-36){$2^+_1$};
				\node at(axis cs:1.5,-38){$4^+_1$};
				\node at(axis cs:2.5,-45.8){$0^+_1$};
				\node at(axis cs:2.5,-43.6){$2^+_1$};
				\node at(axis cs:2.5,-47.9){$4^+_1$};
    \nextgroupplot[
      ymin=-97,ymax=-83,height=3.75cm,
      hide x axis=false,
      axis x line*=bottom,
      xtick pos=bottom,
      xtick={0,1,2,3},
      xtick style={draw=none},
      xticklabels={
        $0.6V_{\alpha\Omega}^{J=2}$,
        $0.9V_{\alpha\Omega}^{J=2}$,
        $V_{\alpha\Omega}^{J=2}$,
        $1.5V_{\alpha\Omega}^{J=2}$
      }]
    \addplot[mark=-, mark size=0.5cm, semithick, line cap=round,
				draw=none, point meta=explicit symbolic,
				]coordinates
				{
                    (3,-85.5)[$-$85.5]
					(3,-86.3)[$-$86.3]
					(3,-95.1)[$-$95.1]
				};
				\node at(axis cs:3.5,-84.8){$0^+_1$};
				\node at(axis cs:3.5,-87.5){$2^+_1$};
				\node at(axis cs:3.5,-94.5){$4^+_1$};
  \end{groupplot}

  \node[anchor=south west]
    at ([xshift=-6mm,yshift=1mm]group c1r1.north west)
    {$E$ (MeV)};

  \foreach \upper/\lower in {1/2,2/3}{
    \foreach \side in {west,east}{
      \begin{scope}[shift={
        ($(group c1r\upper.south \side)!0.5!
          (group c1r\lower.north \side)$)
      }]
        \draw[thick]
          (-1.5mm,-1.2mm) -- (1.5mm,0mm)
          (-1.5mm,0mm) -- (1.5mm,1.2mm);
      \end{scope}
    }
  }
\end{tikzpicture}
    \caption{Energy spectra of the $0_1^+$, $2_1^+$, and $4_1^+$ states of $\prescript{9}{\Omega}{\mathrm{Be}}$, calculated with the scaled interaction $\lambda V_{\alpha\Omega}^{J=2}$ for $\lambda=0.6$, 0.9, 1.0, and 1.5. Breaks in the vertical axis separate the different energy ranges.}
    \label{fig:EnergyVary}
\end{figure}

\begin{table*}
\caption{Energy decomposition of the $0_1^+$, $2_1^+$, and $4_1^+$ states of $^{9}_{\Omega}\mathrm{Be}$, calculated with the scaled interaction $\lambda V_{\alpha\Omega}^{J=2}$. The remaining columns give the expectation values of the individual energy contributions. $T_{\mathrm{eff}}$ denotes the remaining kinetic-energy contribution after subtracting the explicitly listed $\dfrac{l(l+1)}{2\mu}\dfrac{1}{r_3^2}$ centrifugal contribution; their sum gives the total kinetic-energy expectation value. All energies and energy contributions are expressed in MeV. \label{tab:EnergyVary}}
    \begin{ruledtabular}
    \begin{tabular}{cccccccc}
          Interaction&State&$E$&$T_{\text{eff}}$&$\dfrac{l(l+1)}{2\mu}\dfrac{1}{r_3^2}$&$\lambda V_{\alpha\Omega}^{J=2}(\boldsymbol{r}_1)$&$\lambda V_{\alpha\Omega}^{J=2}(\boldsymbol{r}_2)$&$V_{\alpha\alpha}(\boldsymbol{r}_3)$\\
         \hline
         $0.6V_{\alpha\Omega}^{J=2}$&$0^+$&$-$20.1&18.6&0.0&$-$17.6&$-$17.6&$-$3.5\\
         &$2^+$&$-$16.5&19.4&8.8&$-$18.2&$-$18.2&$-$8.4\\
         &$4^+$&$-$13.4&23.2&45.0&$-$22.2&$-$22.2&$-$37.3\\
         $0.9V_{\alpha\Omega}^{J=2}$&$0^+$&$-$39.4&25.3&0.0&$-$31.2&$-$31.2&$-$2.3\\
         &$2^+$&$-$36.7&27.2&10.5&$-$33.2&$-$33.2&$-$8.1\\
         &$4^+$&$-$37.8&32.0&53.3&$-$39.3&$-$39.3&$-$44.6\\
         $V_{\alpha\Omega}^{J=2}$&$0^+$&$-$46.4&27.4&0.0&$-$36.0&$-$36.0&$-$1.7\\
         &$2^+$&$-$44.3&29.4&11.1&$-$38.5&$-$38.5&$-$7.8\\
         &$4^+$&$-$46.6&34.6&55.7&$-$45.2&$-$45.2&$-$46.5\\
         $1.5V_{\alpha\Omega}^{J=2}$&$0^+$&$-$85.5&36.8&0.0&$-$62.6&$-$62.6&2.8\\
         &$2^+$&$-$86.3&40.0&13.6&$-$67.6&$-$67.6&$-$4.7\\
         &$4^+$&$-$95.1&46.1&66.0&$-$76.7&$-$76.7&$-$53.9\\
    \end{tabular}
    \end{ruledtabular}
\end{table*}
}

As shown in Table~\ref{tab:energy spectra}, the $\alpha\alpha$ distance $R_{\alpha\alpha}$ is drastically reduced compared to $^8$Be: from 5.96, 3.60, 2.91~fm in $^8$Be to approximately 2.94, 2.64, and 2.20~fm for the $0^+_1$, $2^+_1$, and $4^+_1$ states, respectively. This dramatic shrinkage of the nuclear core clearly demonstrates the gluelike role of the $\Omega$ baryon. The $R_{\alpha\Omega}$ values (2.07, 1.83, and 1.61~fm) are comparable to the $\alpha\Omega$ distance in the two-body $\prescript{5}{\Omega}{\text{He}}$ system ($1.35$~fm), indicating a compact three-body structure.

For the $\prescript{5}{\Omega}{\mathrm{He}}$ system, our result is relatively consistent with those reported in Refs.~\cite{Etminan:2019gds,Filikhin:2024hrd,Etminan:2024uvc}, whereas it differs substantially from that of Ref.~\cite{Filikhin:2025ivu}. The relatively strong binding obtained in the present work is also consistent with our result for the $NN\Omega$ system. In the $^{1,6}S_{5/2}$ channel, we obtain a binding energy of $23.5$ MeV for $NN\Omega$, which is comparable in magnitude to previous predictions of $16.34$--$22.0$ MeV in Refs.~\cite{Garcilazo:2018gkb,Garcilazo:2019igo,Zhang:2021vsf,Etminan:2023gzh,Filikhin:2025nvt,Filikhin:2025ivu}, although our result lies toward the more deeply bound side. This overall consistency supports the strong $N\Omega$ attraction generated by the original HAL QCD interaction and, consequently, makes a relatively deeply bound $\prescript{5}{\Omega}{\mathrm{He}}$ system natural within the same interaction framework.

The much smaller binding energy obtained for $\prescript{5}{\Omega}{\mathrm{He}}$ in Ref.~\cite{Filikhin:2025ivu} can be traced mainly to the different $\alpha\Omega$ interaction adopted there. As illustrated in Fig.~\ref{fig:potential-comparison}, Ref.~\cite{Filikhin:2025ivu} employs the double-Gaussian parametrization shown by the black dash-dotted curve, which is fitted to the long-range part of the $\alpha\Omega$ potential, whereas in the present work we use the folded $\alpha\Omega$ potential shown by the red solid curve, obtained directly from the original HAL QCD $N\Omega$ interaction. The latter retains a substantially stronger attraction at short and intermediate distances and therefore produces a considerably larger binding energy for $\prescript{5}{\Omega}{\mathrm{He}}$.

A similar situation occurs for the $\prescript{9}{\Omega}{\mathrm{Be}}$ system. Although our results are qualitatively consistent with those of Ref.~\cite{Filikhin:2025ivu} in predicting a bound $\Omega$-nuclear system, our binding energies are significantly larger. This difference originates from the same choice of the $\alpha\Omega$ interaction: the shallower double-Gaussian potential adopted in Ref.~\cite{Filikhin:2025ivu} produces substantially weaker binding, whereas the folded potential used in our calculation preserves the strong attraction contained in the original HAL QCD $N\Omega$ potential. We consider the direct use of the original HAL QCD interaction in the folding procedure to be physically better motivated, since it avoids an additional phenomenological modification of the resulting $\alpha\Omega$ interaction. The resulting deep binding and small intercluster distances of $\prescript{9}{\Omega}{\mathrm{Be}}$ indicate a pronounced gluelike effect of the $\Omega$ baryon and suggest the possible formation of compact, deeply bound $\Omega$-nuclear clusters. Future experimental studies of such exotic states would therefore provide valuable constraints on the $N\Omega$ interaction.

	\subsection{The $\prescript{5}{\Omega_{ccc}}{\mathrm{He}}$ and $\prescript{9}{\Omega_{ccc}}{\text{Be}}$ systems}

As shown in Fig.~\ref{fig:energy spectra}, the $\Omega_{ccc}$ baryon produces a much weaker gluelike effect than the $\Omega$ baryon, but a stronger one than the $J/\psi$. Similar to the $J/\psi$ case, the $0_1^+$ state of $\prescript{9}{\Omega_{ccc}}{\mathrm{Be}}$ becomes bound, whereas the $2_1^+$ and $4_1^+$ states remain resonant. However, the $\Omega_{ccc}$ baryon shifts the spectrum further downward and generally leads to narrower resonances, reflecting the stronger $\alpha\Omega_{ccc}$ attraction compared with the $\alpha J/\psi$ interaction. Moreover, as the $\alpha\Omega_{ccc}$ attraction increases from the $J=2$ interaction to the spin-averaged and then to the $J=1$ interaction, the spectrum is systematically lowered and the $2_1^+$ and $4_1^+$ resonances become progressively narrower.

To better understand the three-body spectrum, we first consider the simpler $\alpha\Omega_{ccc}$ two-body system, $\prescript{5}{\Omega_{ccc}}{\mathrm{He}}$. As listed in Table~\ref{tab:energy spectra}, all three $\alpha\Omega_{ccc}$ interactions support weakly bound states. For $R_\alpha=1.70$ fm, the binding energies are $0.571$, $0.939$, and $1.70$ MeV for $V_{\alpha\Omega_{ccc}}^{J=2}(r)$, $V_{\alpha\Omega_{ccc}}^{\rm av}(r)$, and $V_{\alpha\Omega_{ccc}}^{J=1}(r)$, respectively. The corresponding rms distances are $R_{\alpha\Omega_{ccc}}=3.67$, $3.05$, and $2.48$ fm. Thus, the $\prescript{5}{\Omega_{ccc}}{\mathrm{He}}$ system is only weakly bound and has a relatively extended spatial configuration. The increase in binding energy accompanied by the decrease in $R_{\alpha\Omega_{ccc}}$ also reflects the increasing attractive strength from the $J=2$ interaction to the spin-averaged and $J=1$ interactions.

Our results for $\prescript{5}{\Omega_{ccc}}{\mathrm{He}}$ are considerably less bound than those reported in Ref.~\cite{Filikhin:2025ivu}. In that work, binding energies of $3.53$ and $9.90$ MeV were obtained using two different parametrizations of the $\Omega_{ccc}N$ interaction. The origin of this quantitative discrepancy differs from that in the $\Omega$ sector. In Ref.~\cite{Filikhin:2025ivu}, the $N\Omega_{ccc}$ interaction was not taken directly from a HAL QCD calculation. Instead, it was constructed by scaling a modified $N\Omega$ interaction according to
\begin{equation}
\widetilde{V}_{N\Omega_{ccc}}(r)=c_2\,\widetilde{V}_{N\Omega}(c_1 r),
\end{equation}
where the parameters $c_1$ and $c_2$ were introduced based on an approximate analogy between the $\Omega\Omega$ and $\Omega_{ccc}\Omega_{ccc}$ interactions. The resulting $N\Omega_{ccc}$ interaction was then used to construct an effective $\alpha\Omega_{ccc}$ potential. In the present work, by contrast, we directly employ the $N\Omega_{ccc}$ potentials obtained by the HAL QCD Collaboration in both the $J=1$ and $J=2$ channels and construct the corresponding $\alpha\Omega_{ccc}$ interactions through the folding procedure. The substantially different two-body inputs therefore naturally lead to sizable differences in the predicted binding energies. 

The structural properties of $\prescript{9}{\Omega_{ccc}}{\mathrm{Be}}$ also reflect the relatively weak gluelike effect of the $\Omega_{ccc}$ baryon. For the $0_1^+$ state, the $\alpha\alpha$ rms distance is reduced from $5.96$ fm in $^8$Be to about $4$ fm, indicating a noticeable contraction of the diffuse $^8$Be core. In contrast, the $\alpha\alpha$ distances of the $2_1^+$ and $4_1^+$ resonances remain comparable to, or slightly larger than, those of the corresponding states in $^8$Be. Thus, the gluelike effect of the $\Omega_{ccc}$ baryon is much weaker than that of the $\Omega$ baryon and is most clearly manifested in the ground state.

Our binding energies for $\prescript{9}{\Omega_{ccc}}{\mathrm{Be}}$ are substantially smaller than those reported in Ref.~\cite{Filikhin:2025ivu}. This difference has the same origin as in the $\prescript{5}{\Omega_{ccc}}{\mathrm{He}}$ system: Ref.~\cite{Filikhin:2025ivu} employs a modeled $N\Omega_{ccc}$ interaction constructed from the modified $N\Omega$ potential, whereas we directly use the $N\Omega_{ccc}$ potentials obtained by HAL QCD. The resulting difference in the underlying two-body interaction naturally leads to different $\alpha\Omega_{ccc}$ folded potentials and, consequently, different binding energies.

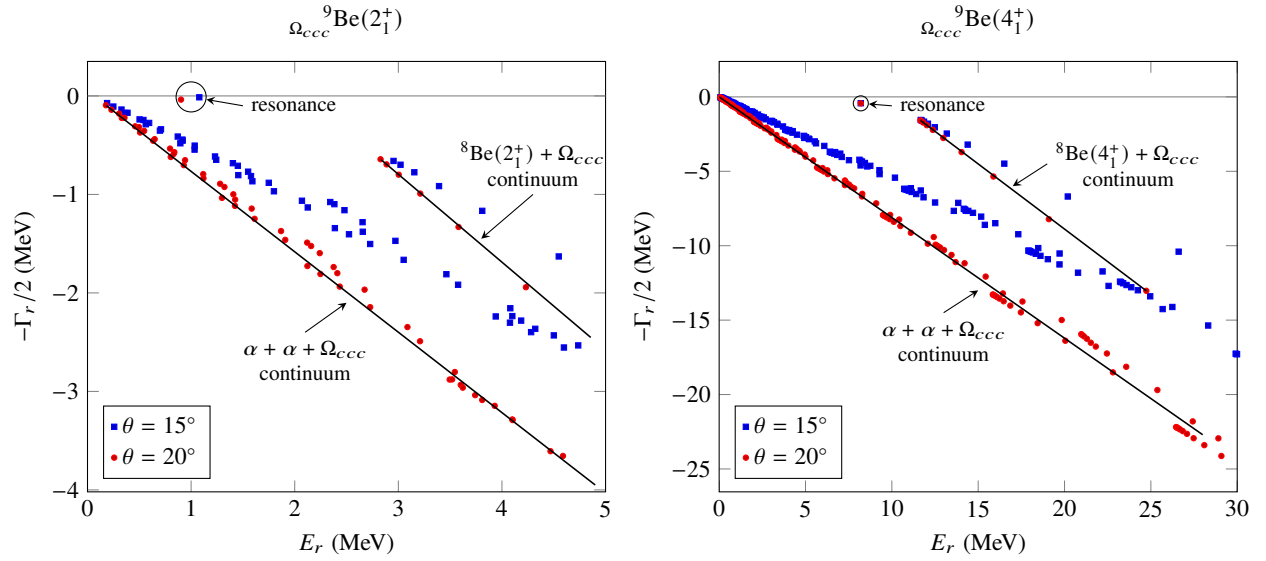
\begin{figure*}[htbp]
		\begin{tikzpicture}
			\begin{axis}[
				title=$\prescript{9}{\Omega_{ccc}}{\text{Be}}(2_1^+)$,
				xlabel=$E_r$ (MeV),
				xmin=0, xmax=5, 
				ylabel=$-\Gamma_r/2$ (MeV),
				legend entries={$\theta=15^\circ$,$\theta=20^\circ$},
				legend pos=south west,
				mark size=1pt,
				set layers,
				]
				\addplot+[only marks,mark=square*]coordinates{
					(0.189066,-0.0727065)
(0.24949,-0.108035)
(0.327953,-0.138382)
(0.362518,-0.173045)
(0.385961,-0.171237)
(0.50261,-0.238498)
(0.539666,-0.244197)
(0.561646,-0.287972)
(0.594775,-0.274251)
(0.698968,-0.352308)
(0.709129,-0.339212)
(0.870842,-0.412584)
(0.896181,-0.481033)
(0.916336,-0.443475)
(0.917872,-0.45404)
(1.07917,-0.0141458)
(1.03432,-0.504538)
(1.03351,-0.545047)
(1.23087,-0.614838)
(1.24237,-0.648169)
(1.40331,-0.682909)
(1.4509,-0.7092)
(1.45552,-0.804221)
(1.55014,-0.769131)
(1.57794,-0.813841)
(1.59139,-0.866652)
(1.74949,-0.882046)
(1.79927,-0.96823)
(2.0697,-1.06405)
(2.12563,-1.13249)
(2.3429,-1.07734)
(2.38283,-1.0989)
(2.47999,-1.16)
(2.38745,-1.3425)
(2.52382,-1.40485)
(2.65469,-1.28177)
(2.6567,-1.379)
(2.9536,-0.660027)
(3.02108,-0.698564)
(2.72807,-1.50339)
(3.15444,-0.77526)
(2.97171,-1.47145)
(3.05127,-1.66421)
(3.3928,-0.91558)
(3.46381,-1.81036)
(3.80994,-1.1667)
(3.57674,-1.91695)
(3.93991,-2.23899)
(4.0807,-2.1557)
(4.10018,-2.23536)
(4.07737,-2.30225)
(4.18529,-2.28112)
(4.54816,-1.62969)
(4.28136,-2.39821)
(4.3215,-2.36429)
(4.50075,-2.43137)
(4.59822,-2.55425)
(4.73622,-2.53296)
(5.09539,-2.8132)
				};
				
				\addplot+[only marks,mark=*]coordinates{
(0.17793,-0.0947987)
(0.231201,-0.140169)
(0.305055,-0.179771)
(0.331263,-0.223761)
(0.356631,-0.222087)
(0.459933,-0.30881)
(0.497044,-0.316331)
(0.507428,-0.371527)
(0.546491,-0.355286)
(0.633236,-0.454909)
(0.648106,-0.439048)
(0.902454,-0.0374502)
(0.796554,-0.534617)
(0.83824,-0.571392)
(0.803347,-0.619741)
(0.831284,-0.589469)
(0.941552,-0.652482)
(0.929443,-0.702662)
(1.1168,-0.794052)
(1.11921,-0.83597)
(1.27992,-0.892021)
(1.32149,-0.924372)
(1.29804,-1.03518)
(1.40759,-0.99904)
(1.42509,-1.05242)
(1.42307,-1.11618)
(1.58451,-1.14387)
(1.61251,-1.24762)
(1.86799,-1.37193)
(1.90853,-1.46114)
(2.12001,-1.48957)
(2.15808,-1.5239)
(2.12228,-1.72714)
(2.24141,-1.59627)
(2.24798,-1.80821)
(2.82822,-0.642912)
(2.37544,-1.73737)
(2.8865,-0.695142)
(2.41118,-1.79966)
(3.00355,-0.79944)
(2.43456,-1.93591)
(2.67419,-1.96697)
(3.21179,-0.991735)
(2.72808,-2.1448)
(3.58106,-1.33076)
(3.08791,-2.34658)
(3.20954,-2.49068)
(3.54511,-2.80302)
(3.49544,-2.87964)
(3.5192,-2.87834)
(3.60231,-2.93358)
(4.2312,-1.94224)
(3.62196,-2.9618)
(3.74146,-3.03814)
(3.80938,-3.08696)
(3.93003,-3.14628)
(4.10025,-3.28472)
(4.10266,-3.29084)
(4.46858,-3.60705)
(4.58893,-3.65552)
(5.39056,-2.9914)
				};

				\draw[gray,thin](axis cs:0,0)--(axis cs:5,0);
				
				\begin{pgfonlayer}{axis foreground}
					\draw[semithick](axis cs:2.83577,-0.649315)--(axis cs:4.8567,-2.45194);
					\draw[-stealth](axis cs:4.3,-1)node[above,align=center,font=\footnotesize]{$\prescript{8}{}{\text{Be}}(2^+_1)+\Omega_{ccc}$\\continuum}--(axis cs:3.8,-1.45);
					\draw[semithick](axis cs:0.183459,-0.099623)--(axis cs:4.9,-3.95);
					\draw[-stealth](axis cs:2.1,-2.4)node[below,align=center,font=\footnotesize]{$\alpha+\alpha+\Omega_{ccc}$\\continuum}--(axis cs:2.5,-2.1);
					\node[circle,draw,inner sep=4](res)at(axis cs:1,-0.01){};
					\draw[-stealth](axis cs:1.5,-0.1)node[right]{\footnotesize resonance}to(res);
				\end{pgfonlayer}
			\end{axis}
		\end{tikzpicture}
		\begin{tikzpicture}
			\begin{axis}[
				title=$\prescript{9}{\Omega_{ccc}}{\text{Be}}(4_1^+)$,
				xlabel=$E_r$ (MeV),
				xmin=0, xmax=30, 
				ylabel=$-\Gamma_r/2$ (MeV),
				legend entries={$\theta=15^\circ$,$\theta=20^\circ$},
				legend pos=south west,
				mark size=1pt,
				set layers,
				]

				\addplot+[only marks,mark=square*]coordinates{
(0.0712791,-0.0294034)
(0.103724,-0.0438336)
(0.130302,-0.0634794)
(0.149931,-0.0651631)
(0.162713,-0.0778745)
(0.208955,-0.0993286)
(0.219429,-0.0986013)
(0.244609,-0.129475)
(0.2794,-0.133582)
(0.277038,-0.1439)
(0.326874,-0.152135)
(0.323537,-0.165538)
(0.386775,-0.186892)
(0.393038,-0.198899)
(0.443201,-0.244133)
(0.475657,-0.258592)
(0.49568,-0.238703)
(0.500934,-0.253036)
(0.522254,-0.28031)
(0.55402,-0.272063)
(0.591729,-0.313704)
(0.670296,-0.339627)
(0.699183,-0.367043)
(0.764155,-0.379718)
(0.777766,-0.437295)
(0.819977,-0.410927)
(0.810243,-0.451779)
(0.866831,-0.453034)
(0.856884,-0.47355)
(0.935806,-0.477887)
(0.926706,-0.507208)
(1.03331,-0.559914)
(1.13623,-0.593495)
(1.19489,-0.610427)
(1.20171,-0.645759)
(1.24774,-0.639058)
(1.36056,-0.703569)
(1.33659,-0.759935)
(1.36908,-0.774434)
(1.41576,-0.796257)
(1.46621,-0.78398)
(1.48577,-0.830096)
(1.56194,-0.81936)
(1.5929,-0.883185)
(1.7592,-0.967322)
(1.88936,-0.986658)
(1.90003,-1.01209)
(1.94081,-1.01661)
(2.02576,-1.10599)
(2.0491,-1.07619)
(2.24667,-1.18862)
(2.26808,-1.29773)
(2.30057,-1.31223)
(2.34728,-1.33409)
(2.41741,-1.36808)
(2.44791,-1.32996)
(2.52499,-1.42157)
(2.58445,-1.38087)
(2.69198,-1.50618)
(2.95461,-1.64113)
(3.02587,-1.6076)
(3.06827,-1.63101)
(3.14656,-1.69362)
(3.17366,-1.70355)
(3.36482,-1.80055)
(3.37979,-1.86578)
(3.69675,-1.98686)
(3.82026,-2.19388)
(3.85275,-2.20839)
(3.89946,-2.23025)
(3.96968,-2.26434)
(4.05981,-2.2297)
(4.07757,-2.31822)
(4.25669,-2.30119)
(4.24553,-2.40361)
(4.50881,-2.53917)
(4.89747,-2.61831)
(4.9288,-2.65132)
(4.92771,-2.75744)
(5.02785,-2.70552)
(5.18489,-2.8055)
(5.22309,-2.84357)
(5.53761,-2.99233)
(5.60946,-3.12061)
(6.08989,-3.29608)
(6.40706,-3.68737)
(6.43956,-3.70188)
(6.48628,-3.72376)
(6.55651,-3.75788)
(6.71535,-3.70919)
(6.66464,-3.812)
(6.83333,-3.89828)
(7.01451,-3.81306)
(7.09854,-4.03531)
(8.19862,-0.425835)
(7.51701,-4.2529)
(8.12168,-4.22518)
(8.17871,-4.29347)
(8.27027,-4.34355)
(8.18835,-4.60495)
(8.44055,-4.44192)
(8.57657,-4.67406)
(8.75155,-4.63069)
(9.28291,-4.92245)
(9.28967,-5.19229)
(10.1801,-5.42693)
(11.6698,-1.5502)
(11.7345,-1.5886)
(11.8621,-1.66473)
(12.0898,-1.80164)
(10.7193,-6.17704)
(10.7518,-6.19156)
(10.7985,-6.21342)
(10.8688,-6.24759)
(12.4877,-2.04262)
(10.9769,-6.3018)
(11.1011,-6.13701)
(11.1462,-6.38866)
(11.4132,-6.52763)
(11.6735,-6.28911)
(13.1806,-2.46518)
(11.8354,-6.74775)
(12.5041,-7.09659)
(14.3915,-3.20638)
(13.8461,-7.12214)
(13.5874,-7.65885)
(14.1682,-7.51939)
(14.2362,-7.56668)
(14.2864,-7.64967)
(14.4262,-7.66079)
(14.6983,-7.79111)
(16.5129,-4.48548)
(15.1737,-8.04558)
(15.3857,-8.59871)
(15.9911,-8.48758)
(17.3225,-9.23007)
(17.9106,-10.3289)
(17.9431,-10.3434)
(17.9898,-10.3653)
(18.0601,-10.3995)
(18.1684,-10.4538)
(18.48,-10.1605)
(18.3376,-10.5409)
(20.1881,-6.69387)
(18.6059,-10.6813)
(19.0323,-10.9054)
(19.714,-10.5238)
(19.7081,-11.2572)
(20.7825,-11.8128)
(22.2027,-11.7283)
(22.5509,-12.7016)
(23.2143,-12.4131)
(23.2737,-12.4503)
(23.3941,-12.5233)
(23.592,-12.6411)
(23.8656,-12.7919)
(24.247,-12.9932)
(24.9619,-13.4005)
(26.6083,-10.4001)
(25.6814,-14.2579)
(26.2452,-14.1165)
(28.3185,-15.3622)
(29.9101,-17.2568)
(29.9426,-17.2714)
(29.9893,-17.2932)
(30.0596,-17.3274)
				};
				
				\addplot+[only marks,mark=*]coordinates{
(0.0664286,-0.0381789)
(0.096375,-0.0568761)
(0.118638,-0.0819864)
(0.138811,-0.0844916)
(0.14856,-0.10064)
(0.190979,-0.128396)
(0.20223,-0.127716)
(0.219748,-0.166829)
(0.255134,-0.172644)
(0.249678,-0.185517)
(0.299739,-0.196832)
(0.292312,-0.213512)
(0.352659,-0.241502)
(0.355766,-0.256643)
(0.395412,-0.31423)
(0.425358,-0.332959)
(0.452229,-0.308516)
(0.453521,-0.326502)
(0.468073,-0.361053)
(0.503948,-0.351429)
(0.531483,-0.404223)
(0.606652,-0.438279)
(0.629089,-0.473128)
(0.693757,-0.490296)
(0.691352,-0.562554)
(0.743535,-0.530515)
(0.721312,-0.581314)
(0.780485,-0.584043)
(0.764051,-0.609471)
(0.845978,-0.616619)
(0.827747,-0.652978)
(0.924645,-0.721076)
(1.0234,-0.76546)
(1.07999,-0.787501)
(1.07698,-0.831915)
(1.12745,-0.824462)
(1.22751,-0.907509)
(1.18567,-0.977332)
(1.21563,-0.996108)
(1.2584,-1.02433)
(1.31517,-1.01014)
(1.32222,-1.06806)
(1.40593,-1.05659)
(1.41957,-1.13665)
(1.57022,-1.24534)
(1.70225,-1.27265)
(1.70584,-1.30548)
(1.74667,-1.30992)
(1.8106,-1.4243)
(1.84378,-1.38743)
(2.01931,-1.53237)
(2.00961,-1.66871)
(2.03958,-1.68749)
(2.08236,-1.71575)
(2.14627,-1.75966)
(2.1899,-1.7132)
(2.24394,-1.82877)
(2.3195,-1.78028)
(2.39521,-1.93804)
(2.63259,-2.11262)
(2.71817,-2.07457)
(2.75537,-2.10282)
(2.82588,-2.18671)
(2.83926,-2.19236)
(3.01792,-2.32133)
(3.01522,-2.40227)
(3.31386,-2.56222)
(3.38259,-2.82077)
(3.41256,-2.83956)
(3.45534,-2.86784)
(3.5193,-2.91186)
(3.62558,-2.87277)
(3.61718,-2.98142)
(3.81309,-2.96821)
(3.76919,-3.09171)
(4.00717,-3.26686)
(4.39333,-3.3976)
(4.40987,-3.43152)
(4.3849,-3.54916)
(4.49943,-3.50282)
(4.63906,-3.62606)
(4.67006,-3.67439)
(4.95245,-3.8742)
(4.99795,-4.01771)
(5.44529,-4.26849)
(5.67076,-4.74077)
(5.70073,-4.75956)
(5.74352,-4.78784)
(5.80746,-4.83192)
(5.99243,-4.78336)
(5.9055,-4.90176)
(6.058,-5.01314)
(6.26737,-4.93335)
(6.29755,-5.19014)
(8.20247,-0.435086)
(6.67548,-5.47151)
(7.28248,-5.61465)
(7.30064,-5.72242)
(7.38275,-5.78302)
(7.28101,-5.9277)
(7.53333,-5.90313)
(7.6601,-6.05705)
(7.81891,-6.1463)
(8.28777,-6.51432)
(8.27103,-6.68905)
(9.09402,-7.15317)
(11.6577,-1.57802)
(11.7147,-1.62759)
(11.8266,-1.72606)
(12.0255,-1.90321)
(9.48515,-7.94143)
(9.51513,-7.96022)
(9.5579,-7.98851)
(9.62193,-8.03261)
(12.3729,-2.21517)
(9.7199,-8.1026)
(9.90181,-7.92929)
(9.87276,-8.2146)
(10.1136,-8.39414)
(12.9784,-2.76027)
(10.4252,-8.24042)
(10.4947,-8.67872)
(11.0987,-9.13041)
(14.0386,-3.71086)
(12.4259,-9.41859)
(12.0782,-9.86356)
(12.5309,-9.93129)
(12.5943,-9.97906)
(12.6983,-10.0575)
(12.8001,-10.1263)
(13.021,-10.292)
(15.8825,-5.35012)
(13.4555,-10.6164)
(13.6927,-11.1003)
(14.2119,-11.1791)
(15.4334,-12.0745)
(15.8462,-13.279)
(15.8762,-13.2978)
(15.919,-13.3261)
(19.0926,-8.21652)
(15.9829,-13.3702)
(16.0811,-13.4402)
(16.4302,-13.2082)
(16.234,-13.5527)
(16.4754,-13.7336)
(16.8598,-14.0236)
(17.5644,-13.7493)
(17.4708,-14.4799)
(18.4432,-15.2042)
(19.8356,-14.9925)
(20.0459,-16.3843)
(20.9689,-15.9431)
(21.0172,-15.9912)
(21.1193,-16.0853)
(21.2934,-16.2497)
(21.5204,-16.5171)
(21.8104,-16.7756)
(24.7369,-13.0293)
(22.4527,-17.242)
(22.8034,-18.5087)
(23.5729,-18.1343)
(25.3769,-19.6968)
(26.4604,-22.1854)
(26.4904,-22.2041)
(26.5331,-22.2325)
(26.5972,-22.2765)
(26.6951,-22.3466)
(26.8482,-22.4592)
(27.4242,-21.8036)
(27.0901,-22.6413)
(27.4758,-22.9349)
(28.0947,-23.4016)
(28.91,-22.9427)
(29.0834,-24.128)
(30.6734,-25.296)
				};
				
				\draw[gray,thin](axis cs:0,0)--(axis cs:30,0);
				
				\begin{pgfonlayer}{axis foreground}
					\draw[semithick](axis cs:11.6643,-1.58421)--(axis cs:24.7369,-13.0293);
					\draw[-stealth](axis cs:19,-4.5)node[right,align=center,font=\footnotesize]{$\prescript{8}{}{\text{Be}}(4^+_1)+\Omega_{ccc}$\\continuum}--(axis cs:17,-6);
					\draw[semithick](axis cs:0,0)--(axis cs:28,-22.7);
					\draw[-stealth](axis cs:13,-14.5)node[below,align=center,font=\footnotesize]{$\alpha+\alpha+\Omega_{ccc}$\\continuum}--(axis cs:15,-13);
					\node[circle,draw,inner sep=2](res)at(axis cs:8.20247,-0.435086){};
					\draw[-stealth](axis cs:10,-0.6)node[right]{\footnotesize resonance}to(res);
				\end{pgfonlayer}
			\end{axis}
		\end{tikzpicture}
		\caption{Complex-energy spectra of the $2_1^+$ and $4_1^+$ states of $\prescript{9}{\Omega_{ccc}}{\text{Be}}$, calculated with the spin-averaged $V_{\alpha\Omega_{ccc}}^{\rm av}(r)$ interaction for $R_\alpha=1.70$ fm at the scaling angles $\theta=15^\circ$ and $20^\circ$. The eigenvalues associated with the $\alpha+\alpha+\Omega_{ccc}$ and $^{8}\mathrm{Be}(J^+)+\Omega_{ccc}$ continua rotate with $\theta$, whereas the resonance poles remain stationary in the complex-energy plane. The complex energy is expressed as $E=E_r-i\Gamma_r/2$. \label{fig:OmegacccBe}}
	\end{figure*}

Finally, as shown in Fig.~\ref{fig:OmegacccBe}, the $2_1^+$ and $4_1^+$ states are identified as resonances and lie close to the corresponding $^8\mathrm{Be}(2_1^+)+\Omega_{ccc}$ and $^8\mathrm{Be}(4_1^+)+\Omega_{ccc}$ thresholds, respectively. The attractive $\alpha\Omega_{ccc}$ interaction slightly lowers these resonant states relative to the corresponding excitations of the $^8\mathrm{Be}$ core, providing another manifestation of the gluelike effect of the $\Omega_{ccc}$ baryon. However, this effect is not strong enough to pull the $2_1^+$ and $4_1^+$ states below threshold, so they remain resonant rather than becoming bound states.
	
	\section{SUMMARY}\label{sec4}
In this work, we have systematically investigated the bound and resonant states of the $\alpha+\alpha+\Omega$ and $\alpha+\alpha+\Omega_{ccc}$ three-body systems within the Gaussian expansion method and the complex scaling method. The $N\Omega$ and $N\Omega_{ccc}$ interactions obtained from lattice QCD calculations by the HAL QCD Collaboration were employed as the basic two-body inputs. The corresponding $\alpha\Omega$ and $\alpha\Omega_{ccc}$ interactions were constructed through a single-folding procedure using the nucleon density distribution of the $\alpha$ particle. To estimate the uncertainty associated with the nuclear size, calculations were performed for three representative $\alpha$-particle rms matter radii, $R_\alpha=1.84$, $1.70$, and $1.56$ fm. In addition, the $N\Omega/\Omega_{ccc}$ and $NN\Omega/\Omega_{ccc}$ systems were examined as benchmarks for understanding the underlying two-body interactions.

For the $\Omega$ sector, the strong attraction predicted by the HAL QCD $N\Omega$ interaction leads to pronounced binding already at the few-baryon level. We obtain a shallow $N\Omega$ bound state and a considerably more compact $NN\Omega$ state with a binding energy of $23.5$ MeV. After embedding the $\Omega$ baryon into the $\alpha\alpha$ system, the attraction is further enhanced. For $R_\alpha=1.70$ fm, the $0_1^+$, $2_1^+$, and $4_1^+$ states of $^{9}_{\Omega}\mathrm{Be}$ are found at approximately $-46.4$, $-44.3$, and $-46.6$ MeV, respectively. The three states are therefore nearly degenerate, and an unconventional inversion between the $0_1^+$ and $4_1^+$ levels is predicted. Meanwhile, the corresponding $\alpha\alpha$ rms distances decrease from $5.96$, $3.60$, and $2.91$ fm in $^{8}\mathrm{Be}$ to about $2.94$, $2.64$, and $2.20$ fm, respectively. Such a substantial shrinkage of the nuclear core demonstrates that the $\Omega$ baryon plays an exceptionally strong gluelike role and may favor the formation of compact, deeply bound $\Omega$-nuclear clusters.

The situation is qualitatively different in the $\Omega_{ccc}$ sector. The HAL QCD $N\Omega_{ccc}$ interaction does not support bound $N\Omega_{ccc}$ or $NN\Omega_{ccc}$ states, whereas the folded $\alpha\Omega_{ccc}$ interactions are sufficiently attractive to produce weakly bound $\prescript{5}{\Omega_{ccc}}{\mathrm{He}}$ states. For $\prescript{9}{\Omega_{ccc}}{\text{Be}}$, all three $\alpha\Omega_{ccc}$ interactions considered in this work produce a bound $0_1^+$ state, while the $2_1^+$ and $4_1^+$ states remain resonant. With increasing attraction from the $J=2$ interaction to the spin-averaged interaction and finally to the $J=1$ interaction, the spectrum shifts systematically toward lower energies and the widths of the excited resonances decrease. For example, at $R_\alpha=1.70$ fm, the spin-averaged interaction gives $E=-2.40$ MeV for the $0_1^+$ state, while the $2_1^+$ and $4_1^+$ resonances appear at $E_r=1.08$ and $8.20$ MeV with widths of $0.028$ and $0.85$ MeV, respectively.

The spatial structures further emphasize the different roles played by the two baryons. While the $\Omega$ baryon strongly compresses the $\alpha\alpha$ core for all three low-lying states, the gluelike effect of the $\Omega_{ccc}$ baryon is considerably weaker and is most evident in the $0_1^+$ ground state, for which the $\alpha\alpha$ rms distance is reduced from $5.96$ fm in $^{8}\mathrm{Be}$ to approximately $4$ fm. The different spectra and structural behaviors of $^{9}_{\Omega}\mathrm{Be}$ and $\prescript{9}{\Omega_{ccc}}{\text{Be}}$ directly reflect the substantial difference between the underlying $N\Omega$ and $N\Omega_{ccc}$ interactions. Our results therefore provide useful theoretical predictions for future searches for $\Omega$- and $\Omega_{ccc}$-containing nuclei and may offer a complementary way to constrain the corresponding baryon-nucleon interactions.

\vfil

\begin{acknowledgments}

This work is supported by the Natural Science Foundation of Gansu Province (No. 26RCKA012 and No. 25JRRA799), the National Natural Science Foundation of China under Grants No. 12335001 and No. 12247101, the ``111 Center" under Grant No. B20063, the fundamental Research Funds for the Central Universities (lzujbky-2023-stlt01), and Lanzhou City High-Level Talent Funding.

\medskip
\noindent\textbf{AI Usage Declaration:} During the preparation of this work, the authors used ChatGPT to assist with literature searches, language editing, and figure preparation. The research ideas, physical framework, code writing for numerical calculations, and analysis procedures were developed by the authors. All content prepared with AI assistance was reviewed and revised by the authors, who take full responsibility for the accuracy of the results and the contents of this paper. 

\end{acknowledgments} 

\section*{DATA AVAILABILITY}
The data that support the findings of this article are openly available~\cite{ParticleDataGroup:2024,Mohr:2024kco}.


	\bibliography{References}
	
\end{document}